\pdfoutput=1
\documentclass{article}

\usepackage{PRIMEarxiv}

\usepackage[utf8]{inputenc}
\usepackage[T1]{fontenc}
\usepackage{amsmath}
\usepackage{booktabs}
\usepackage{tabularx}
\usepackage{longtable}
\usepackage{capt-of}
\usepackage{float}
\usepackage{microtype}
\usepackage{graphicx}
\usepackage{fancyhdr}
\usepackage{xspace}
\xspaceaddexceptions{\%}
\usepackage[round]{natbib}
\usepackage{url}
\usepackage{xurl}
\usepackage{hyperref}
\hypersetup{colorlinks=true, linkcolor=blue, citecolor=blue, urlcolor=blue}
\hypersetup{pdftitle={AI Isn't Getting Dumber, We're Getting Pickier: Rising Standards in Human Evaluation of Large Language Models},
  pdfauthor={Zihan Chen, Lei Nico Zheng, Di Zhu},
  pdfkeywords={human evaluation, large language models, Chatbot Arena, LLM leaderboards, evaluation drift, expectation confirmation, adaptation level, reference points, human feedback}}
\graphicspath{{figures/}}

\newtheorem{hypothesis}{Hypothesis}

\newcommand{\NRaw}{1{,}799{,}991\xspace}
\newcommand{\NWindow}{1{,}795{,}852\xspace}
\newcommand{\NFrozen}{1{,}260{,}967\xspace}
\newcommand{\NSnapTwo}{430{,}370\xspace}
\newcommand{\NModels}{129\xspace}
\newcommand{\NSnapModels}{33\xspace}
\newcommand{\NOpenModels}{73\xspace}
\newcommand{\NAliasModels}{23\xspace}
\newcommand{\NFrozenModels}{106\xspace}
\newcommand{\NPairsFrozen}{2{,}231\xspace}
\newcommand{\BothBadAll}{17.0\xspace}
\newcommand{\TieAll}{18.7\xspace}
\newcommand{\ShareFrozen}{70\xspace}
\newcommand{\EloFirst}{823\xspace}
\newcommand{\EloLast}{1067\xspace}
\newcommand{\EloGain}{245\xspace}
\newcommand{\ExpScore}{78\xspace}
\newcommand{\ShareAboveBest}{80\xspace}
\newcommand{\CorrEloNoTies}{.999\xspace}
\newcommand{\CorrMeanEloMonth}{.99\xspace}
\newcommand{\BBAllFirst}{19.4\xspace}
\newcommand{\BBAllLast}{20.6\xspace}
\newcommand{\BBAllMin}{13.8\xspace}
\newcommand{\BBAllMax}{20.6\xspace}
\newcommand{\MainB}{$+$1.71\xspace}
\newcommand{\MainAbs}{1.71\xspace}
\newcommand{\MainSE}{0.18\xspace}
\newcommand{\MainMean}{17.1\xspace}
\newcommand{\MainSEPair}{0.07\xspace}
\newcommand{\MainYear}{21\xspace}
\newcommand{\MonthLastB}{17.8\xspace}
\newcommand{\MonthLastSE}{2.7\xspace}
\newcommand{\BalEarly}{15.5\xspace}
\newcommand{\BalLate}{23.3\xspace}
\newcommand{\BalDiff}{7.8\xspace}
\newcommand{\BalLo}{6.4\xspace}
\newcommand{\BalHi}{9.1\xspace}
\newcommand{\BalGap}{5.3\xspace}
\newcommand{\BalPerMonth}{1.48\xspace}
\newcommand{\TieB}{$-$0.09\xspace}
\newcommand{\TieSE}{0.22\xspace}
\newcommand{\DecisiveAbs}{1.62\xspace}
\newcommand{\DecisiveSE}{0.14\xspace}
\newcommand{\StrongerB}{$+$0.13\xspace}
\newcommand{\StrongerSE}{0.13\xspace}
\newcommand{\StrongerMean}{66.7\xspace}
\newcommand{\CtrlPromptB}{$+$1.63\xspace}
\newcommand{\CtrlConvB}{$+$1.57\xspace}
\newcommand{\CtrlRespB}{$+$1.59\xspace}
\newcommand{\SingleTurnB}{$+$1.60\xspace}
\newcommand{\EasyB}{$+$1.30\xspace}
\newcommand{\EasySE}{0.17\xspace}
\newcommand{\HighFreqB}{$+$1.61\xspace}
\newcommand{\HighFreqSE}{0.96\xspace}
\newcommand{\EnglishB}{$+$1.53\xspace}
\newcommand{\SnapB}{$+$1.55\xspace}
\newcommand{\DocB}{$+$1.41\xspace}
\newcommand{\DocSE}{0.20\xspace}
\newcommand{\OpenWB}{$+$2.12\xspace}
\newcommand{\StableB}{$+$1.16\xspace}
\newcommand{\StableSE}{0.16\xspace}
\newcommand{\StableSnapB}{$+$0.97\xspace}
\newcommand{\UISegB}{$+$2.61\xspace}
\newcommand{\UIAllSegB}{$+$1.03\xspace}
\newcommand{\UIAllSegSE}{0.31\xspace}
\newcommand{\UIAllExclB}{$+$1.77\xspace}
\newcommand{\TierWeakB}{$+$1.63\xspace}
\newcommand{\TierMidB}{$+$1.76\xspace}
\newcommand{\TierStrongB}{$+$1.75\xspace}
\newcommand{\MainTwentyFourAbs}{1.91\xspace}
\newcommand{\MainTwentyFourSE}{0.23\xspace}
\newcommand{\StrongTwentyFourB}{0.00\xspace}
\newcommand{\StrongTwentyFourSE}{0.13\xspace}
\newcommand{\StrongNoTiesB}{$+$0.09\xspace}
\newcommand{\StrongCFB}{$+$0.13\xspace}
\newcommand{\StrongCFTwentyFourB}{$+$0.01\xspace}
\newcommand{\SpanThreeB}{$+$1.54\xspace}
\newcommand{\SpanSixB}{$+$1.10\xspace}
\newcommand{\CellsB}{$+$1.45\xspace}
\newcommand{\NoProlificB}{$+$1.65\xspace}
\newcommand{\NoSurgeB}{$+$1.56\xspace}
\newcommand{\SpanTwentyFour}{7.4\xspace}
\newcommand{\MainTwentyFourSpan}{14\xspace}
\newcommand{\StrongTwentyFourSpanLo}{$-$1.9\xspace}
\newcommand{\StrongTwentyFourSpanHi}{1.9\xspace}
\newcommand{\NWeeksTwentyFour}{33\xspace}
\newcommand{\NWeeks}{68\xspace}
\newcommand{\StrongLateLo}{4.1\xspace}
\newcommand{\StrongLateHi}{6.0\xspace}
\newcommand{\StrongMinAbs}{9.4\xspace}
\newcommand{\BBDecMonth}{6.5\xspace}
\newcommand{\BBLateRise}{11\xspace}
\newcommand{\RobMin}{$+$0.97\xspace}
\newcommand{\RobMax}{$+$2.61\xspace}
\newcommand{\RobN}{28\xspace}
\newcommand{\RobNotSig}{one\xspace}
\newcommand{\OsterDeltaRec}{36\xspace}
\newcommand{\OsterBStarTen}{$+$0.98\xspace}
\newcommand{\OsterRmaxZero}{0.23\xspace}
\newcommand{\OsterRmaxZeroRatio}{twelve\xspace}
\newcommand{\OsterRZero}{0.001\xspace}
\newcommand{\OsterROne}{0.020\xspace}
\newcommand{\ShareTwentyFour}{92\xspace}
\newcommand{\MedSpanVotes}{2.1\xspace}
\newcommand{\ProlificShare}{22\xspace}
\newcommand{\SurgeBefore}{5{,}808\xspace}
\newcommand{\SurgeAfter}{7{,}511\xspace}
\newcommand{\NCommits}{56\xspace}
\newcommand{\NMajor}{eleven\xspace}
\newcommand{\UIAllExclShare}{36\xspace}
\newcommand{\UIMeanAbs}{1.6\xspace}
\newcommand{\PlaceboMeanAbs}{1.6\xspace}
\newcommand{\PlaceboN}{36\xspace}
\newcommand{\PlaceboShareSig}{14\xspace}
\newcommand{\UIShareSig}{18\xspace}
\newcommand{\FrontGPTFourO}{5.0\xspace}
\newcommand{\FrontTurbo}{1.3\xspace}
\newcommand{\NDocModels}{18\xspace}
\newcommand{\NStableModels}{8\xspace}
\newcommand{\NServingChecked}{40\xspace}
\newcommand{\UISigNeg}{two\xspace}
\newcommand{\NLang}{seven\xspace}
\newcommand{\LangMin}{$+$1.52\xspace}
\newcommand{\LangMax}{$+$3.75\xspace}
\newcommand{\NStableModelsWord}{eight\xspace}
\newcommand{\BalPairsWord}{fourteen\xspace}
\newcommand{\WJB}{$+$1.90\xspace}
\newcommand{\WJAbs}{1.90\xspace}
\newcommand{\WJSE}{0.13\xspace}
\newcommand{\WJJudges}{24{,}508\xspace}
\newcommand{\WJThreeAbs}{1.74\xspace}
\newcommand{\NewB}{$+$1.25\xspace}
\newcommand{\NewSE}{0.13\xspace}
\newcommand{\NewCtrlB}{$+$1.15\xspace}
\newcommand{\NewEngB}{$+$1.17\xspace}
\newcommand{\VetAbs}{2.56\xspace}
\newcommand{\VetSE}{0.31\xspace}
\newcommand{\PrevB}{$-$0.11\xspace}
\newcommand{\PrevSE}{0.09\xspace}
\newcommand{\PrevJB}{$-$0.25\xspace}
\newcommand{\PrevJSE}{0.09\xspace}
\newcommand{\PlatMeanAbs}{7.32\xspace}
\newcommand{\JPB}{$+$2.30\xspace}
\newcommand{\JPSE}{0.68\xspace}
\newcommand{\JPN}{11{,}642\xspace}
\newcommand{\JPGroups}{3{,}437\xspace}
\newcommand{\JPJudges}{1{,}292\xspace}
\newcommand{\JPPairs}{459\xspace}
\newcommand{\JPCtrlB}{$+$2.24\xspace}
\newcommand{\RAModels}{eighteen\xspace}
\newcommand{\RAMeanRaw}{$+$1.40\xspace}
\newcommand{\RAMeanAdj}{$+$0.19\xspace}
\newcommand{\RARawPos}{Seventeen\xspace}
\newcommand{\RARawPosLow}{seventeen\xspace}
\newcommand{\RAAdjPos}{three\xspace}
\newcommand{\RAAdjNeg}{one\xspace}
\newcommand{\WJLkB}{$+$1.50\xspace}
\newcommand{\WJNoProlificB}{$+$1.83\xspace}
\newcommand{\WJClJWSE}{0.15\xspace}
\newcommand{\WJShareIds}{6\xspace}
\newcommand{\WJShareVotes}{29\xspace}
\newcommand{\NewNoSurgeB}{$+$1.17\xspace}
\newcommand{\NewMonthLast}{11.6\xspace}
\newcommand{\NewMonthLastSE}{2.7\xspace}

\title{AI Isn't Getting Dumber, We're Getting Pickier:\\Rising Standards in Human Evaluation\\of Large Language Models}

\author{
  Zihan Chen \\
  Stevens Institute of Technology \\
  Hoboken, NJ, USA \\
  \texttt{zchen61@stevens.edu} \\
  \And
  Lei Nico Zheng \\
  University of Massachusetts Boston \\
  Boston, MA, USA \\
  \texttt{lei.zheng@umb.edu} \\
  \And
  Di Zhu \\
  Stevens Institute of Technology \\
  Hoboken, NJ, USA \\
  \texttt{dzhu1@stevens.edu} \\
}

\begin{document}
\maketitle

\begin{abstract}
Users often complain that AI assistants such as ChatGPT are getting worse. Some versions do change, but human evaluation leaves a second possibility untested: the people judging may change. We separate the two on Chatbot Arena, a public platform where people compare answers from two anonymous large language models (LLMs) and pick the better one or say both are bad. We study \NFrozen blind votes cast from May 2023 to August 2024 on pairs of models whose versions are fixed: dated vendor snapshots and models with fixed open weights. For the same pair of unchanged models, the share of votes saying ``both are bad'' rises by \MainAbs percentage points a month on a mean of \MainMean\%, and by \MainTwentyFourAbs points a month in 2024. In \BalPairsWord pairs observed often in both 2023 and 2024, it rose in every one, from \BalEarly\% to \BalLate\%. The ranking does not trend: in 2024, the stronger model's share of decisive votes is flat. The rise holds within the same voter and on the same pair, and across arrival cohorts from their first recorded vote. It holds with prompt controls, in every major language, for dated vendor snapshots, open-weight models and models with stable answer length, and between interface changes. The drift is shared across models: month effects estimated from other models' pairs remove it from each model's own pairs. Platform-wide, the average model would be expected to score \ExpScore\% against the 2023 average, yet the share of ``both are bad'' votes ended near where it began. The pattern fits a standard that follows the category's progress, so a product that stands still is judged as declining. Absolute verdicts on AI drifted while relative comparisons held steady, an evaluation drift that matters for LLM leaderboards and for the human feedback used to train models.
\end{abstract}


\section{Introduction}
\label{sec:intro}

Complaints that an AI assistant ``got dumber'' have followed almost every widely used large language model (LLM). Users say a model has become lazy, that it refuses more, that its answers have lost their edge. These complaints matter. They feed public debate about whether AI progress has stalled, they shape whether people keep paying for a service, and they are often read as evidence that a provider quietly changed its product. Sometimes the product did change. \citet{chen2024chatgpt} show that the March and June 2023 versions of the model behind ChatGPT behaved differently on several tasks, so a model offered under the same name can change underneath its users.

A complaint, however, is a comparison between what a person receives and what the person expected. If expectations rise, an answer that once satisfied will disappoint even if nothing about it has changed. This comparison lies at the core of the expectation-confirmation account of satisfaction with information systems \citep{oliver1980cognitive, bhattacherjee2001understanding}, and adaptation-level theory in psychology makes the same point: judgments are made against a reference that drifts toward recent experience \citep{helson1964adaptation, brickman1971hedonic, frederick1999hedonic}. Generative AI is an extreme case for such drift. Better models arrived every few months between 2023 and 2024, and many people used several of them. If each new model raised the bar, people would judge yesterday's AI more harshly even when yesterday's AI had not changed at all.

The two explanations, a declining product and rising standards, predict the same complaints, and in ordinary use they are hard to separate. The model behind a product name can change, the people complaining change, and the questions people ask change as they learn what AI can do. Separating them requires the same unchanged product, judged repeatedly over a long period, by people who do not know which product they are judging, with each verdict recorded as it is made.

The distinction matters for decisions that organizations make from human evaluations of AI. Providers watch user ratings for signs that a release has regressed and decide what to change. Buyers choose models from public leaderboards built on human votes. Developers train models on human judgments of which answers are acceptable. Each of these decisions reads a fall in human ratings as information about the system. If the evaluators' standard moves, the same data mislead: a stable product looks like a regressing one, and labels collected in different periods encode different standards.

Existing theory offers competing answers to where such a decline comes from. Expectation-confirmation research explains evaluations through expectations that users form and revise by using the focal system \citep{bhattacherjee2001understanding, bhattacherjee2004understanding}. Research on online ratings attributes the decline of a fixed product's ratings to changes in who rates \citep{li2008self, godes2012sequential}. Adaptation-level theory predicts instead that each rater's reference rises with experience across a category \citep{helson1964adaptation}. These accounts locate the decline in different places, and they have not been compared in a setting that holds the system fixed, keeps the evaluation blind and follows the same raters over time. We ask three research questions. First, do human evaluations of unchanged AI systems become harsher over time? Second, does the change sit in absolute or in relative judgments, and in the raters themselves or in the composition of the rater pool? Third, can an evaluation system remove the change, and how?

Chatbot Arena provides such a setting \citep{zheng2023judging, chiang2024arena}. On Arena, a person writes a prompt, receives answers from two anonymous models side by side, and votes: the left answer is better, the right answer is better, it is a tie, or both are bad. Model names are revealed only after the vote. Many of the models Arena served were fixed versions: dated snapshots that vendors publish as unchanging (such as \texttt{gpt-4-0314}) and models whose open weights are fixed. Pairs of unchanged models therefore met the public blind, again and again, and the overlapping pairs span more than a year. We study the \NFrozen votes cast on such pairs from May 2023 to August 2024 and ask whether verdicts on the same unchanged models changed over time. Holding the pair of models fixed, a change in verdicts must come from the people judging, from the prompts they write, or from the platform. We test each of these in turn.

The answer to the first question is that people became steadily harsher. We report six findings, which answer the three questions in turn.
\begin{enumerate}
  \item \textbf{Same models, harsher verdicts.} Within the same pair of frozen models, the share of votes saying ``both are bad'' rises by \MainAbs percentage points a month (SE \MainSE) on a mean of \MainMean\%, and by \MainTwentyFourAbs points a month in 2024, when the data are densest. In the \BalPairsWord pairs observed often in both 2023 and 2024, the share rose in every pair, from \BalEarly\% to \BalLate\%.
  \item \textbf{The bar moves, not the ranking.} The extra ``both are bad'' votes come out of decisive votes; ties do not change. In 2024, when most votes were cast, the stronger model's share of decisive votes shows no trend (\StrongTwentyFourB points a month, SE \StrongTwentyFourSE) while ``both are bad'' rises by \MainTwentyFourAbs. What changes is how often neither answer is acceptable, not which answer is preferred.
  \item \textbf{Not the prompts, the models or the interface.} The rise holds with controls for prompt language, content, difficulty tags and conversation, in single-turn conversations, on easy English prompts and in each of the \NLang most common languages. It holds for pairs of dated OpenAI and Anthropic snapshots, of open-weight models and of models with stable answer length, and within the periods between changes to Arena's interface. These measured changes do not account for it.
  \item \textbf{People become pickier.} The rise appears within the same voter over time (\WJB points a month among \WJJudges judge identifiers seen over at least a month) and within the same voter on the same pair (\JPB, with identifier-by-pair fixed effects). It also appears across arrival cohorts: a voter arriving in 2024 judges the same frozen pair more harshly on their first recorded vote than a voter who arrived in 2023 (\NewB points per month of arrival date).
  \item \textbf{No momentary contrast.} The strength of the models a voter happened to see in their previous battle, which the platform draws for them, does not make their next verdict harsher.
  \item \textbf{A shared drift that frozen references remove.} \RARawPos of \RAModels frequently judged frozen models show a significant rise in ``both are bad'' on their own pairs. Month effects estimated only from other models' pairs remove it: the average trend falls from \RAMeanRaw to \RAMeanAdj points a month.
\end{enumerate}
On the platform as a whole, the pattern resembles a treadmill. A model drawn from the August 2024 battles would be expected to score about \ExpScore\% against one drawn from May 2023, yet the share of ``both are bad'' votes across all battles was \BBAllFirst\% in May 2023 and \BBAllLast\% in August 2024.

These results make three contributions. First, they show that evaluations of an information system can decline while the system stays fixed, through a change located in the evaluators. Continuance research models expectations about a system and how they change with use of that system \citep{bhattacherjee2001understanding, bhattacherjee2004understanding, kim2005longitudinal}. We observe evaluations of systems that never changed becoming harsher as the rest of their category improved, within the same voters and for newcomers from their first recorded vote. Adaptation-level theory explains this pattern with a reference point formed across the category rather than by the focal system \citep{helson1964adaptation}.

Second, the results identify a rater-side mechanism behind declining ratings. Online ratings for the same product tend to fall over time, and prior work attributes the decline to changes in who rates: early buyers like a product more than later ones, and later reviewers react to earlier reviews \citep{li2008self, godes2012sequential, moe2012online}. Because Arena holds the product fixed and lets us follow the same voter, it shows that the same voter becomes harsher, so even a stable pool of raters would produce declining scores for an unchanged product.

Third, the results inform the design of evaluation systems. Comparative judgments are known to be more stable than absolute ones \citep{thurstone1927law, parducci1965category, stewart2005absolute}, and rating scales drift in their meaning \citep{garg2021designing, filippas2022reputation}. Human-preference platforms and the human feedback used to train LLMs mix both kinds of judgment \citep{christiano2017deep, ouyang2022training, chiang2024arena}. On Arena, the absolute part drifted at an annualized rate of about \MainYear points while the relative part showed no trend in 2024, so an absolute label collected in 2023 does not mean what the same label means in 2024. Because the drift is shared across models, frozen reference pairs judged in the same period can re-anchor absolute verdicts, and we show that this works out of sample.

For practice, the results caution against reading complaints about AI decline at face value: an AI product that stands still will be judged as getting worse. For evaluation platforms, absolute verdicts should be time-stamped and re-anchored against frozen references, and comparisons across periods should rest on relative judgments or on fixed reference items.

The rest of the paper proceeds as follows. Section~\ref{sec:background} develops the framework and hypotheses. Section~\ref{sec:data} describes Arena and the data, Section~\ref{sec:approach} the empirical approach, and Section~\ref{sec:results} the results. Section~\ref{sec:discussion} discusses implications, limitations and future work, and Section~\ref{sec:conclusion} concludes.

\section{Background and Hypotheses}
\label{sec:background}

\subsection{Does the model change, or does the judge?}

When users say an AI system has become worse, two kinds of change could be responsible. The first is a change in the system. Hosted LLMs are updated behind stable product names, and \citet{chen2024chatgpt} document large differences in behaviour between two versions of the same named service released three months apart. The second is a change in the person. The same output can be judged differently if the standard the person applies has moved.

Evidence on the first kind of change is growing. Evidence on the second is scarce, because most data on how people judge AI come from a single moment, from changing systems, or from people who know which system they are using. Human-preference data are nonetheless central to how AI is built and compared. Preference comparisons train models through reinforcement learning from human feedback \citep{christiano2017deep, ouyang2022training}, and public leaderboards rank models by aggregating crowdsourced comparisons \citep{zheng2023judging, chiang2024arena}. Recent work shows that these rankings are sensitive to which models are sampled and how \citep{singh2025leaderboard}, and that dropping a handful of preferences can change which model ranks first \citep{huang2026dropping}. \citet{singh2025leaderboard} also argue that, as prompts shift and models are retired, historical pairwise comparisons cease to represent present-day performance. Their concern is a changing mix of prompts and models. Whether the judges themselves change, so that the same comparison would receive a different verdict later, has not been examined.

\subsection{Reference-dependent evaluation of information systems}

Our framework is reference-dependent evaluation: a person judges an outcome against a reference, and the reference moves with experience. In information systems research, the best-known version is expectation-confirmation theory. Satisfaction depends on how performance compares with prior expectations \citep{oliver1980cognitive}, and users whose expectations are confirmed keep using a system \citep{bhattacherjee2001understanding}. Longitudinal work shows that beliefs and attitudes about a system change after adoption as users gain experience with it \citep{bhattacherjee2004understanding, kim2005longitudinal}, and the comparison is asymmetric: quality that falls short of expectations weighs more heavily than quality that exceeds them \citep{anderson1993antecedents}. Empirical continuance studies measure expectations about the focal system and follow how they change with use of that system. Whether progress elsewhere in a product category moves the standard applied to a fixed system has rarely been tested with field data.

Adaptation-level theory supplies the mechanism for such a category effect. A judgment of any stimulus is made relative to an adaptation level, a reference that pools the stimuli a person has experienced \citep{helson1964adaptation}. As experienced stimuli improve, the reference rises and the same stimulus is judged worse. \citet{brickman1971hedonic} called the aggregate consequence a hedonic treadmill: improvements raise the reference point, so evaluations return toward where they were \citep{frederick1999hedonic, diener2006beyond}. A well-known aggregate pattern consistent with this account is that rising national income has not brought lasting increases in reported happiness \citep{easterlin1974growth}. Range-frequency theory makes a related prediction for category ratings: the label a stimulus receives depends on where it sits among the stimuli a judge has encountered \citep{parducci1965category}. The reference, in these accounts, is formed by everything a person has experienced in the category, not only by the focal object.

Generative AI is a strong test of this category-level account. Between 2023 and 2024 new models arrived every few months, each better than the last, and many people used several assistants. If the reference is set by the category, a model that never changed would be judged more harshly as the models around it improved.

\subsection{A simple model of a verdict}

To make the predictions precise, we separate three things: the vote that is observed, the quality of each answer as the voter perceives it, and the voter's acceptance threshold. Consider a voter who receives two answers with perceived quality $q_a$ and $q_b$ and who holds a threshold $r$, the quality an answer must reach before the voter accepts it. The voter rejects both answers when neither reaches the threshold, and otherwise prefers the better one:
\begin{equation}
\label{eq:vote}
\textit{BothBad} = \mathbf{1}\{\max(q_a, q_b) < r\}, \qquad \textit{A wins} = \mathbf{1}\{q_a > q_b\} \text{ when the vote is decisive}.
\end{equation}
Perceived quality depends on the answer, on the task, and on how closely the voter scrutinizes the answer. Holding the two models fixed, the share of ``both are bad'' votes can therefore rise for three reasons. The threshold $r$ can rise. The voter can scrutinize more closely and notice more flaws in the same answer, which lowers $q$. Or the answers can become worse for reasons outside the voter: prompts that are harder for these models, changes in how the models are served, or changes in the interface. The first two are changes in the judge, and either one makes the voter harder to satisfy; we call a voter who becomes harder to satisfy in either way \emph{pickier}, and refer to the judge-side change as a rising standard of evaluation. The tests in Section~\ref{sec:approach} ask whether the third set of reasons explains the rise.

\begin{hypothesis}[Rising bar]
\label{h:rise}
The share of ``both are bad'' verdicts on the same unchanged pair of AI models rises over time.
\end{hypothesis}

A comparative verdict depends on the difference between two answers seen together, while an absolute verdict depends on a reference held in mind \citep{thurstone1927law}. People make absolute judgments by comparison with the stimuli they have recently encountered \citep{stewart2005absolute}, which is why absolute ratings are sensitive to context while paired comparisons are less so. Rating systems show the consequence. On an online labor market, average feedback scores rose sharply over time without a matching rise in quality, so the same score came to carry less information \citep{filippas2022reputation}, and re-anchoring what each level of a rating scale means can counter such drift \citep{garg2021designing}.

Equation~\eqref{eq:vote} alone does not fix how often the stronger model wins the decisive votes. A higher threshold changes which comparisons end decisively, so the stronger model's share of the remaining decisive votes stays the same only if the comparisons that drop out are ordered like those that remain. The prediction for relative verdicts comes instead from the theory of comparative judgment: choosing the better of two answers seen side by side needs no reference held in mind, so a reference that drifts should leave it largely unchanged. Two alternatives predict that the share falls. If voters became less attentive or less able to tell answers apart, the stronger model's share would move toward one half. If the stronger model of a pair became worse while the other did not, its share would fall.

\begin{hypothesis}[Stable ranking]
\label{h:relative}
Among decisive verdicts on the same unchanged pair, the share won by the stronger model remains stable over time.
\end{hypothesis}

\subsection{Who becomes pickier}

Declining ratings for a fixed product can arise from changes in who rates. Early adopters of a product tend to like it more than later buyers \citep{li2008self}, later reviewers differ from earlier ones \citep{godes2012sequential}, and the decision to post depends on the ratings already posted \citep{moe2012online}. A changing pool of raters could likewise produce harsher verdicts on Arena if, for example, early users were enthusiasts and later users were sceptics. Adaptation predicts change within raters as well: the same person becomes harsher as better AI becomes familiar, and people who arrive later bring a higher standard with them, because better AI was already part of their experience before they arrived. The two accounts can operate together, but they make different predictions. A changing pool predicts no change within a rater, while adaptation predicts change both within raters and across arrival cohorts.

\begin{hypothesis}[Pickier judges]
\label{h:judges}
(a) Verdicts on the same unchanged models become harsher within the same judge over time. (b) Judges who arrive later are harsher on the same models from their first recorded vote.
\end{hypothesis}

Finally, adaptation can be fast or slow. A momentary contrast effect would make a person harsher right after seeing a strong model. Adaptation-level theory describes a reference that pools experience over time \citep{helson1964adaptation}, under which one more encounter shifts the reference only slightly. We therefore expect no sizeable harsher verdict right after an encounter with a strong model.

\begin{hypothesis}[No momentary contrast]
\label{h:gradual}
A judge's verdict on an unchanged pair does not become harsher right after an encounter with a stronger model.
\end{hypothesis}

\section{Setting and Data}
\label{sec:data}

\subsection{Chatbot Arena}

Chatbot Arena is a public website, run by the research organization LMSYS during our period, where anyone can compare LLMs for free \citep{zheng2023judging, chiang2024arena}. In its main mode, a visitor writes a prompt and receives answers from two models drawn by the platform, shown side by side as ``Model A'' and ``Model B''. The visitor may continue the conversation and then votes with one of four buttons: A is better, B is better, tie, or both are bad. The names of the two models are revealed only after the vote. The platform, not the visitor, chooses which two models appear, and it changes its sampling weights as models are added and retired. The votes feed a public leaderboard that is widely cited as a measure of LLM quality.

Three features make Arena suited to our question. Verdicts are blind, so a person cannot judge a model by its reputation. The same models are judged repeatedly over many months by a large and changing public. And many of the models are fixed versions, so the object being judged does not change.

\subsection{Data}

We use the public battle log that LMSYS released to let others reproduce its leaderboard, in its release of 14 August 2024.\footnote{Available at \url{https://storage.googleapis.com/arena_external_data/public/clean_battle_20240814_public.json}; the file we use has SHA-256 checksum \texttt{747c1c93\ldots bd1b}.} It records \NRaw battles from 24 April 2023 to 14 August 2024. Each record gives the two models, the vote, a hashed identifier for the voter, a timestamp, the language of the prompt, the number of turns, token counts for the prompt and both answers, flags for code, math and refusals, and seven prompt-difficulty tags (specificity, domain knowledge, complexity, problem solving, creativity, technical accuracy, real-world application). LMSYS assigned the seven tags after the fact, by having one LLM annotator label every prompt against the same written criteria \citep{li2024hard}, so a tag means the same thing for a prompt written in 2023 as for one written in 2024. The log contains no prompt or answer text.

The voter identifier is derived from the network address from which the vote was cast. LMSYS identifies users by IP address in its analysis of the platform \citep[Section 5.1]{chiang2024arena}, and the code that prepares the battle log builds the voter field from it \citep{fastchat}. We call it a judge identifier and use it to follow the same voter over time.

The ``both are bad'' button was added to the Arena interface on 29 April 2023. We therefore start on 1 May 2023, which leaves \NWindow blind battles. The analysis sample is the \NFrozen of these that pair two frozen models (Section~\ref{sec:frozen}). Table~\ref{tab:sample} summarizes both.

\begin{table}[t]
\centering
\caption{Sample}
\label{tab:sample}
\begin{tabular}{lr}
\toprule
Blind battles, 1 May 2023 to 14 Aug 2024 & 1,795,852 \\
\quad between two frozen models & 1,260,967 \\
\quad between two dated API snapshots & 430,370 \\
\quad between two dated OpenAI or Anthropic snapshots & 220,514 \\
\quad between two fixed open-weight models & 268,423 \\
Models & 129 \\
\quad dated API snapshots & 33 \\
\quad fixed open weights & 73 \\
\quad aliases that may change (excluded) & 23 \\
Frozen model pairs & 2,231 \\
Judge identifiers (hashed) & 491,867 \\
\quad judging at least one frozen pair & 398,059 \\
Share of votes ``both are bad'' (all battles) & 17.0\% \\
Share of votes ``tie'' (all battles) & 18.7\% \\
\bottomrule
\end{tabular}

\par\smallskip
\begin{minipage}{0.92\linewidth}\footnotesize
\textit{Notes.} Blind battles in the public Chatbot Arena battle log after the ``both are bad'' button was added. Frozen models are dated vendor snapshots and models with fixed open weights (Section~\ref{sec:frozen}); aliases whose underlying model can change are excluded from frozen pairs. Judge identifiers are hashed in the public log.
\end{minipage}
\end{table}

\subsection{Frozen models}
\label{sec:frozen}

The design uses models whose versions are fixed. We assign each of the \NModels models in the window to one of three classes by an explicit list (Appendix Table~\ref{tab:models}). The first class, \emph{dated API snapshots} (\NSnapModels models), contains names that vendors publish as fixed versions, such as \texttt{gpt-4-0314}, \texttt{gpt-4-0613}, \texttt{claude-2.1} and \texttt{claude-3-opus-20240229}. Its strictest tier is the \NDocModels dated snapshots from OpenAI and Anthropic, which both vendors offered as pinned versions alongside undated aliases that moved to newer models; the other dated names include preview and experimental releases. The second class, \emph{fixed open weights} (\NOpenModels models), contains models whose weights were published as a single checkpoint, such as \texttt{llama-2-70b-chat} and \texttt{mixtral-8x7b-instruct-v0.1}. We call the models in these two classes \emph{frozen}. The third class, \emph{aliases} (\NAliasModels models), contains names that can point to a changing model: undated vendor aliases (\texttt{claude-1}, \texttt{gemini-pro}), consumer products (\texttt{bard-jan-24-gemini-pro}), models that search the web, and the Vicuna names, which LMSYS upgraded across versions. We exclude the aliases.

A frozen pair is a pair of two frozen models. Frozen pairs account for \NFrozen battles (\ShareFrozen\% of the window), \NPairsFrozen distinct pairs and \NFrozenModels models; \NSnapTwo battles are between two dated API snapshots.

A fixed version can still be served with a different system prompt, decoding setting or limit on answer length. Such changes would usually show up in what the model writes, and the log records the length of every answer. For each frozen model active for at least three months, we estimate the trend in the length of its own answers, holding the opponent, the prompt length and the language fixed (Appendix Table~\ref{tab:serving}). The main snapshots, among them \texttt{gpt-4-0314}, \texttt{gpt-4-0613}, \texttt{gpt-4-1106-preview} and \texttt{claude-2.1}, show small estimated trends; we call the \NStableModelsWord models whose estimated linear trend passes a strict screening rule \emph{length-stable}. Answer length registers serving changes that alter how much a model writes, such as a new system prompt or output limit. We use the dated OpenAI and Anthropic snapshots, the open-weight models and the length-stable models as three separate tests of the serving explanation.

\subsection{Outcomes}
\label{sec:outcomes}

The main outcome is an indicator that a vote is ``both are bad''. In terms of Equation~\eqref{eq:vote}, it records that neither answer, as the voter perceived it, reached the voter's threshold. We also study the share of votes for one of the two answers (decisive votes), the share of ties, and, among decisive votes, whether the stronger of the two models won. The stronger model is defined by a Bradley--Terry rating, which infers each model's strength from its wins and losses against all opponents, estimated once on the whole window \citep{bradley1952rank, hunter2004mm}; a single rating summarizes a frozen model because its version does not change. Across all battles in the window, \BothBadAll\% of votes are ``both are bad'' and \TieAll\% are ties.

Because the vote records the outcome of a comparison, Section~\ref{sec:approach} tests whether measured changes in prompts, models and interface account for the rise in ``both are bad'' votes; the part they do not account for we attribute to the judges.

\section{Empirical Approach}
\label{sec:approach}

\subsection{Within-pair comparison}

The estimand is the change over time in the probability that a vote on the same pair of frozen models is ``both are bad''. The unit of analysis is a vote $i$ on a frozen pair $p$ at time $t$. The main specification is a linear probability model with pair fixed effects:
\begin{equation}
\label{eq:main}
\textit{BothBad}_{ipt} = \alpha_p + \beta \cdot \textit{Months}_t + \mathbf{x}_i'\gamma + \varepsilon_{ipt},
\end{equation}
where $\textit{Months}_t$ counts months since 1 May 2023 and $\mathbf{x}_i$ is an optional vector of prompt characteristics. The pair fixed effect $\alpha_p$ absorbs everything about the two models, including how good they are and how often their answers fall short. The coefficient $\beta$ is therefore identified only from how verdicts on the same two models change over time. Hypothesis~\ref{h:rise} predicts $\beta > 0$. To trace the path rather than impose a line, we also replace $\textit{Months}_t$ with a dummy for each calendar month, with May 2023 as the reference.

Pairs enter and leave the platform at different times, so $\beta$ is estimated from many overlapping pairs rather than from one panel observed throughout. With pair fixed effects, $\beta$ is a weighted average of the slopes within each pair, with more weight on pairs judged often and over longer spans. Section~\ref{sec:results} reports how the votes are spread over time and re-estimates $\beta$ with equal weights for pair-month cells and on long-lived pairs.

The same specification with other outcomes tests Hypothesis~\ref{h:relative}: the share of decisive votes, the share of ties, and, among decisive votes on pairs whose ratings differ by at least 20 points, whether the stronger model won. Because the three vote shares sum to one, a rise in ``both are bad'' with flat ties must be matched by a fall in decisive votes; the informative test is whether the stronger model's share of the remaining decisive votes changes.

Standard errors are clustered two ways, by model pair and by calendar week \citep{cameron2011robust}. Clustering by pair allows any correlation among votes on the same models; clustering by week allows shocks that hit all votes in a given week, such as a news event or a temporary change to the site. Clustering by pair alone gives a standard error less than half as large (\MainSEPair instead of \MainSE for the main estimate), so common weekly shocks matter, and we cluster by pair and week in all specifications that pool votes across judges. The window has \NWeeks calendar weeks, \NWeeksTwentyFour of them in 2024.

\subsection{What else could lower perceived quality}

Equation~\eqref{eq:vote} implies that, with the pair held fixed, more ``both are bad'' votes require a higher threshold, closer scrutiny, or answers that became worse for reasons outside the judge. We test three such reasons.

\textbf{The prompts.} People may ask harder questions over time, or questions that suit these models less well. We add controls in three blocks, ordered by how far they are from the original request. The first block describes the prompt's content: fixed effects for the 20 most common prompt languages, code and math flags, and the seven difficulty tags. The second adds the conversation (multi-turn, prompt length), which can depend on earlier answers, and the third adds the answers (refusal, answer length), which are outcomes of the models. We estimate the model on single-turn conversations, where the prompt is the whole request, and on easy, single-turn English prompts (no complexity, problem-solving, code or math tag). We also estimate it on the subset of prompts that the log flags as recurring at high frequency. Finally, we use the coefficient-stability approach of \citet{oster2019unobservable} to ask how much selection on unobserved prompt features, relative to the observed ones, would be needed to explain the trend away, letting the share of the variation in votes that a full set of controls could explain vary.

\textbf{The serving.} A fixed version could be served with a different configuration. We re-estimate the model on pairs of dated OpenAI and Anthropic snapshots, on pairs of open-weight models, and on pairs of length-stable models.

\textbf{The interface.} Arena's interface was changed several times. We use the public history of the source file of the anonymous battle page in the software that runs Arena \citep{fastchat}: \NCommits commits changed it during the window, \NMajor of them by at least 50 lines. We re-estimate the model with fixed effects for each period between major changes, and separately for each period between any two changes, so that a jump at a change is absorbed and only the trend within periods identifies $\beta$. We also drop the days around each change, estimate the jump at each major change directly with a local regression that includes pair fixed effects, and compare those jumps with jumps estimated in the same way at placebo dates. These tests use commit dates, the recorded timing of each code change.

\subsection{Judges}

The pool of voters changes over time. For Hypothesis~\ref{h:judges}(a), we add identifier fixed effects among judge identifiers observed over at least one month, so that each voter is compared with themselves. Because additive judge and pair effects assume that a voter's taste for a pair is stable, we also estimate the model with a fixed effect for each identifier-by-pair combination, using the combinations judged more than a month apart, which compares the same identifier on the same pair at different times. Within an identifier, calendar time and experience on the platform rise together; we separate them by controlling for the number of earlier recorded votes. For Hypothesis~\ref{h:judges}(b), we keep only each identifier's first recorded vote, for identifiers whose first vote falls inside the window, and estimate Equation~\eqref{eq:main} with arrival date as the time variable. These are arrival cohorts: they compare people who first voted at different dates, on the same pair, at their first recorded vote.

\subsection{Momentary contrast}

Hypothesis~\ref{h:gradual} concerns momentary contrast. The platform draws the pair of models for each battle, so the strength of the models a voter saw in their previous recorded battle was not chosen by the voter. We regress the verdict on a frozen pair on the rating of the better model in the voter's previous recorded battle, with pair, week and arrival-week fixed effects and fixed effects for the number of earlier battles, and separately with identifier fixed effects.

\subsection{Re-anchoring with frozen references}

If the drift is a shift shared by all frozen pairs, it can be measured on some pairs and removed from others. We test this out of sample. For each frozen model with at least 10,000 votes over at least four months, we estimate calendar-month effects on ``both are bad'' from all frozen pairs that do not involve that model, subtract them from the verdicts on the model's own pairs, and estimate the remaining trend with pair fixed effects. A model whose own pairs drifted in ways the other pairs did not would keep a trend after this adjustment.

\section{Results}
\label{sec:results}

\subsection{Better models, steady verdicts}

\begin{figure}[t]
\centering
\includegraphics[width=\linewidth]{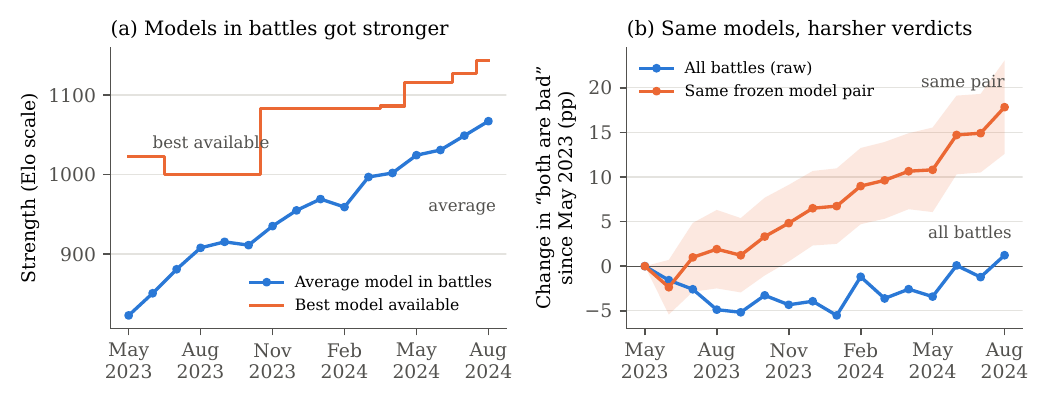}
\caption{Better models, steady verdicts. (a) Average Bradley--Terry rating (Elo scale) of the models in each month's battles, and the rating of the best model with at least 500 battles that month. (b) Change since May 2023 in the share of ``both are bad'' votes: raw, across all battles (blue), and within the same frozen model pair (orange; calendar-month effects from a regression with pair fixed effects; shaded band is the 95\% confidence interval, standard errors clustered by pair and week).}
\label{fig:treadmill}
\end{figure}

Figure~\ref{fig:treadmill} sets the scene. Panel (a) shows how much better the models on Arena became. The average rating of the models in each month's battles rose from \EloFirst in May 2023 to \EloLast in August 2024, a gain of \EloGain points on the Elo scale, the rating scale used in chess, on which a 400-point gap means ten-to-one odds. A model drawn from the August 2024 battles would be expected to score about \ExpScore\% against a model drawn from the May 2023 battles, and \ShareAboveBest\% of the model appearances in August 2024 came from models rated above every model on Arena in May 2023.

Panel (b) shows the verdicts. Across all battles, the share of votes saying ``both are bad'' moved between \BBAllMin\% and \BBAllMax\% from month to month, and it ended the period (\BBAllLast\%) close to where it began (\BBAllFirst\%). Far better models did not reduce the share of answers that people rejected outright. The orange line shows that, within the same pair of frozen models, the share of ``both are bad'' votes rose steadily. The pattern resembles a treadmill: the models improved, and the verdict on a fixed pair of models became harsher. The rest of this section asks whether the harsher verdicts come from the judges.

\subsection{Same models, harsher verdicts}

Table~\ref{tab:main}, panel A, first column, estimates Equation~\eqref{eq:main}. Within the same pair of frozen models, the share of ``both are bad'' votes rises by \MainAbs percentage points a month (SE \MainSE), on a mean of \MainMean\%. The monthly effects in Figure~\ref{fig:treadmill}(b) show a climb rather than a single jump: the share for the same pairs is \BBDecMonth points above its May 2023 level by December 2023 and \MonthLastB points (SE \MonthLastSE) above it by August 2024.

Most of the votes come late in the period. Voters cast \ShareTwentyFour\% of the votes on frozen pairs in 2024, when Arena's traffic was highest, and the vote-weighted median pair was judged over \MedSpanVotes months. The common trend is therefore traced by many overlapping pairs, each observed for part of the period. Panel B of Table~\ref{tab:main} uses only January to August 2024, when the data are densest: the share of ``both are bad'' votes rises by \MainTwentyFourAbs points a month (SE \MainTwentyFourSE). The trend is similar when every pair-month cell gets equal weight (\CellsB) and among pairs judged over at least three months (\SpanThreeB), and smaller but clearly positive among the few pairs judged over at least six months (\SpanSixB), which are mostly 2023-era open models (Table~\ref{tab:robust}).

\begin{figure}[t]
\centering
\includegraphics[width=0.86\linewidth]{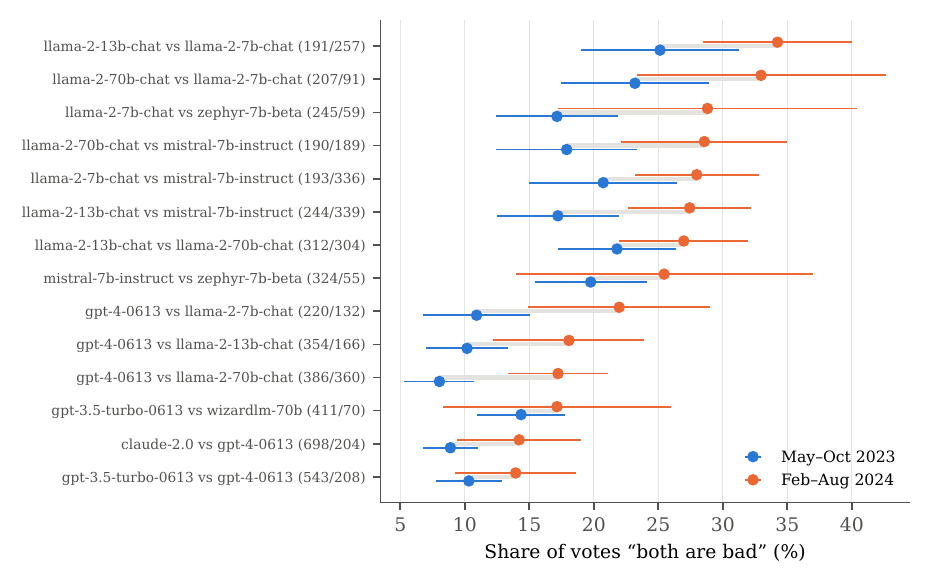}
\caption{Every pair judged often in both periods became ``both are bad'' more often. Frozen model pairs with at least 50 votes in May--October 2023 and in February--August 2024; vote counts in parentheses (early/late). Bars are 95\% binomial confidence intervals.}
\label{fig:pairs}
\end{figure}

Figure~\ref{fig:pairs} shows the pattern without a regression. It plots the \BalPairsWord frozen pairs that were judged at least fifty times both in May--October 2023 and in February--August 2024. The share of ``both are bad'' votes rose in all \BalPairsWord. Weighting each pair by its smaller period count, the share went from \BalEarly\% to \BalLate\%, a rise of \BalDiff points (resampling interval \BalLo to \BalHi). The late votes on these pairs came on average \BalGap months after the early ones, so the rise is about \BalPerMonth points a month. The fourteen pairs confirm the direction of the trend without relying on a regression. The rise appears for \texttt{gpt-4-0613} against \texttt{claude-2.0} and against \texttt{gpt-3.5-turbo-0613}, pairs that were among the strongest available in 2023, as well as for pairs of small open models.

\begin{table}[t]
\centering
\caption{Verdicts on frozen model pairs over time: absolute and relative}
\label{tab:main}
\begin{tabular}{lcccc}
\toprule
 & \multicolumn{3}{c}{Absolute verdict (share of all votes)} & Relative verdict \\
\cmidrule(lr){2-4}\cmidrule(lr){5-5}
 & ``Both are bad'' & One answer wins & Tie & Stronger model wins \\
\midrule
\multicolumn{5}{l}{\textit{A. May 2023 to August 2024}} \\
Change per month (pp) & ${+}1.71^{***}$ & $-1.62^{***}$ & $-0.09$ & ${+}0.13$ \\
 & (0.18) & (0.14) & (0.22) & (0.13) \\
Mean (\%) & 17.1 & 63.9 & 19.0 & 66.7 \\
Votes & 1,260,967 & 1,260,967 & 1,260,967 & 646,299 \\
Model pairs & 2,231 & 2,231 & 2,231 & 1,840 \\
\midrule
\multicolumn{5}{l}{\textit{B. January to August 2024}} \\
Change per month (pp) & ${+}1.91^{***}$ & $-1.96^{***}$ & ${+}0.05$ & $0.00$ \\
 & (0.23) & (0.16) & (0.29) & (0.13) \\
Mean (\%) & 17.2 & 63.5 & 19.3 & 66.5 \\
Votes & 1,155,923 & 1,155,923 & 1,155,923 & 587,377 \\
\bottomrule
\end{tabular}

\par\smallskip
\begin{minipage}{0.95\linewidth}\footnotesize
\textit{Notes.} Linear probability models with model-pair fixed effects; the regressor is months since 1 May 2023. Columns 1--3 use all votes on frozen pairs; the three outcomes sum to one. Column 4 uses decisive votes on pairs whose ratings differ by at least 20 Elo points; the outcome is that the higher-rated model won. Standard errors (in parentheses) are clustered by pair and by calendar week. $^{*}p<0.05$, $^{**}p<0.01$, $^{***}p<0.001$.
\end{minipage}
\end{table}

\subsection{The bar moves, not the ranking}

\begin{figure}[t]
\centering
\includegraphics[width=\linewidth]{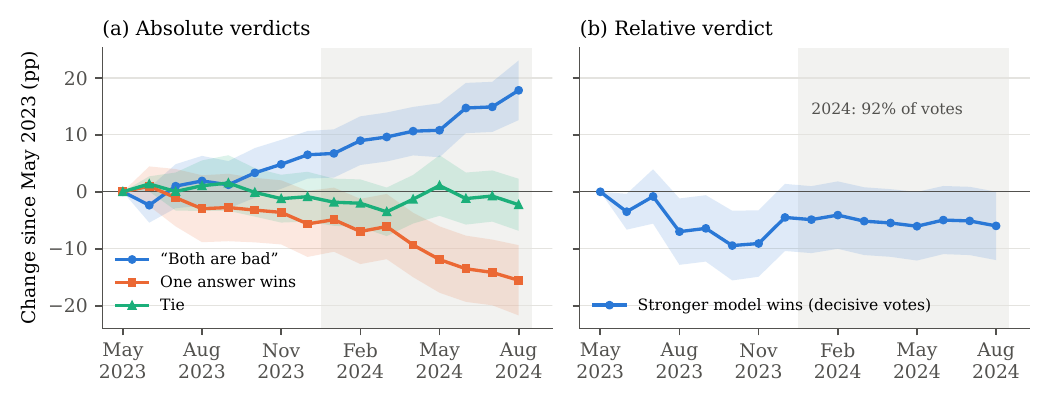}
\caption{Absolute verdicts drift; the relative verdict is flat from late 2023. Calendar-month effects (relative to May 2023) from regressions with model-pair fixed effects. (a) Shares of all votes on frozen pairs. (b) Share of decisive votes won by the higher-rated model, pairs whose ratings differ by at least 20 Elo points. Shaded bands are 95\% confidence intervals, standard errors clustered by pair and week. The grey area marks 2024.}
\label{fig:absrel}
\end{figure}

Table~\ref{tab:main}, columns 2--4, and Figure~\ref{fig:absrel} show where the extra ``both are bad'' votes come from. The share of decisive votes falls by \DecisiveAbs points a month (SE \DecisiveSE), and ties do not change (\TieB points a month, SE \TieSE). In aggregate, the share of ``both are bad'' votes grows at the expense of decisive votes, not of ties.

Among decisive votes on pairs whose ratings differ, the stronger model wins \StrongerMean\% of the time, and its linear trend over the whole period is \StrongerB points a month (SE \StrongerSE). Figure~\ref{fig:absrel}(b) shows that the stronger model's share fell during the first months of the sample, when few votes were cast on a small set of early models, reaching \StrongMinAbs points below its May 2023 level. From December 2023 onward it stays between \StrongLateLo and \StrongLateHi points below that level, while the share of ``both are bad'' votes climbs by about \BBLateRise points over the same months. In 2024 (Table~\ref{tab:main}, panel B), the trend in the stronger model's share is \StrongTwentyFourB points a month (SE \StrongTwentyFourSE). Over the \SpanTwentyFour months of 2024 in the data, the 95\% interval for the change in the stronger model's share runs from \StrongTwentyFourSpanLo to \StrongTwentyFourSpanHi points, while the share of ``both are bad'' votes rose by about \MainTwentyFourSpan points.

This result does not depend on how ``stronger'' is defined. Ratings that drop all ties correlate at \CorrEloNoTies with the main ratings and give a trend of \StrongNoTiesB points a month. Cross-fitted ratings, estimated on alternate calendar weeks and applied to votes in the other weeks, give \StrongCFB over the whole period and \StrongCFTwentyFourB in 2024. In 2024, voters who picked one answer picked the stronger model's answer about as often at the end of the period as at the start, while rejecting both answers far more often, as Hypotheses~\ref{h:rise} and~\ref{h:relative} predict.

\subsection{Not the prompts, the models or the interface}

\begin{figure}[t]
\centering
\includegraphics[width=0.8\linewidth]{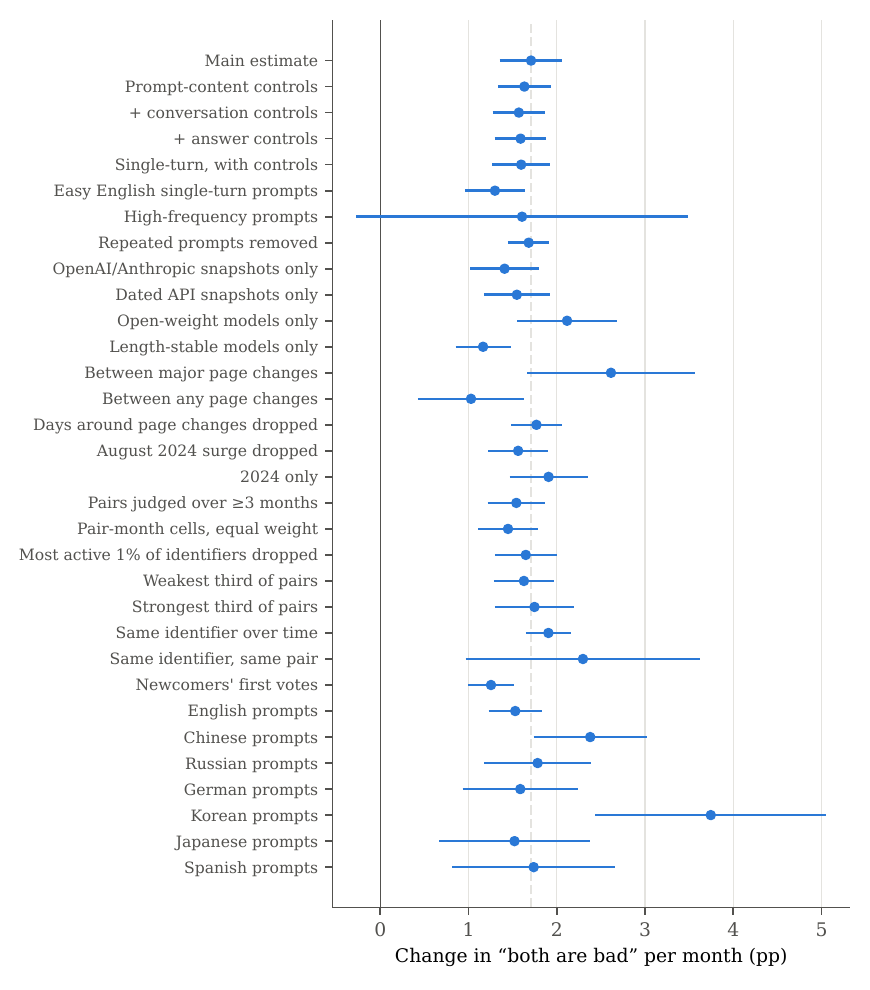}
\caption{One estimate, many cuts. Change in the share of ``both are bad'' votes per month within frozen model pairs, with 95\% confidence intervals. The dashed line marks the main estimate.}
\label{fig:robust}
\end{figure}

Table~\ref{tab:robust} and Figure~\ref{fig:robust} test the explanations outside the judge. Across the \RobN specifications in Table~\ref{tab:robust}, the trend ranges from \RobMin to \RobMax points a month; it is positive in every one and statistically significant in all but \RobNotSig.

\textbf{Prompts.} Controlling for prompt content (language fixed effects, code, math and all seven difficulty tags) gives \CtrlPromptB points a month; adding the conversation gives \CtrlConvB, and adding the answers gives \CtrlRespB. In single-turn conversations with prompt-content controls the trend is \SingleTurnB, and on easy, single-turn English prompts, where the task is simplest and changes least, it is \EasyB (SE \EasySE). It is \EnglishB for English prompts and between \LangMin and \LangMax in each of the \NLang most common prompt languages (Appendix Table~\ref{tab:language}). On the subset of prompts flagged as recurring at high frequency, the point estimate is \HighFreqB with a wide interval (SE \HighFreqSE), because these votes are few. The controls raise the within-pair $R^2$ from \OsterRZero to \OsterROne while moving the coefficient only from \MainB to \CtrlConvB. Under the proportional-selection assumption of \citet{oster2019unobservable}, and with her suggested maximum $R^2$ of 1.3 times the controlled $R^2$, selection on unobserved prompt features would have to be \OsterDeltaRec times as strong as selection on the observed ones to bring the trend to zero (Appendix Table~\ref{tab:oster}). Appendix Table~\ref{tab:oster} varies that maximum. With selection on unobserved features as strong as on observed ones, the trend reaches zero only if that maximum $R^2$ is \OsterRmaxZero, about \OsterRmaxZeroRatio times what all observed prompt features explain; at a maximum of 0.10 the adjusted trend is \OsterBStarTen.

\textbf{Models and serving.} For pairs of two dated OpenAI or Anthropic snapshots, the trend is \DocB points a month (SE \DocSE); for all dated snapshots it is \SnapB, and for pairs of two open-weight models \OpenWB. For pairs of length-stable models it is \StableB (SE \StableSE), and \StableSnapB for length-stable snapshots. These subsets cover different models and periods, and the trend is positive and significant in each.

\textbf{Interface.} With fixed effects for each period between the \NMajor major changes to the battle page, the trend within periods is \UISegB points a month; with fixed effects for each of the periods between all \NCommits changes, most of them a few days long, it is \UIAllSegB (SE \UIAllSegSE). Dropping the three days around every change, \UIAllExclShare\% of the votes, gives \UIAllExclB. Estimated directly with a local regression, the jumps at the major changes average \UIMeanAbs points in absolute value, about the same as at \PlaceboN placebo dates (\PlaceboMeanAbs points), and the \UISigNeg that are significant are negative (Appendix Table~\ref{tab:ui}). On 7 August 2024, in the last week of the log, daily traffic rose from about \SurgeBefore to \SurgeAfter battles and the share of ``both are bad'' votes jumped; dropping that week gives \NoSurgeB. Within the periods between observed changes to the page, the trend stays positive, at \UIAllSegB to \UISegB points a month depending on how finely the periods are drawn.

\textbf{Pair strength and prolific voters.} The trend is similar for the weakest third of pairs (\TierWeakB), the middle third (\TierMidB) and the strongest third (\TierStrongB). The most active 1\% of judge identifiers cast \ProlificShare\% of the votes on frozen pairs; dropping them gives \NoProlificB.

\begin{table}[t]
\centering
\caption{Robustness: change in ``both are bad'' per month within frozen model pairs}
\label{tab:robust}
\small
\begin{tabularx}{\textwidth}{Xrrrr}
\toprule
Specification & pp/month & (SE) & Mean (\%) & Obs. \\
\midrule
Main estimate & ${+}1.71^{***}$ & (0.18) & 17.1 & 1,260,967 \\
\multicolumn{5}{l}{\textit{Prompts}} \\
Prompt content: language fixed effects, code, math, seven difficulty tags & ${+}1.63^{***}$ & (0.15) & 17.1 & 1,260,967 \\
\quad and conversation: multi-turn, prompt length & ${+}1.57^{***}$ & (0.15) & 17.1 & 1,260,967 \\
\quad and answers: refusal, answer length & ${+}1.59^{***}$ & (0.15) & 17.1 & 1,260,967 \\
Single-turn conversations, prompt-content controls and prompt length & ${+}1.60^{***}$ & (0.17) & 17.2 & 1,087,718 \\
Easy, English, single-turn prompts, prompt length & ${+}1.30^{***}$ & (0.17) & 14.7 & 314,486 \\
No refusal flagged, prompt and conversation controls & ${+}1.60^{***}$ & (0.14) & 16.6 & 1,197,586 \\
English prompts only & ${+}1.53^{***}$ & (0.15) & 17.2 & 716,620 \\
Prompts flagged as recurring at high frequency only & ${+}1.61$ & (0.96) & 11.1 & 115,020 \\
High-frequency and unsampled duplicate prompts removed & ${+}1.68^{***}$ & (0.12) & 17.7 & 1,145,947 \\
\multicolumn{5}{l}{\textit{Models}} \\
Both models are dated OpenAI or Anthropic snapshots & ${+}1.41^{***}$ & (0.20) & 13.0 & 220,514 \\
Both models are dated API snapshots & ${+}1.55^{***}$ & (0.19) & 14.0 & 430,370 \\
Both models have fixed open weights & ${+}2.12^{***}$ & (0.29) & 23.2 & 268,423 \\
Both models have stable answer length & ${+}1.16^{***}$ & (0.16) & 11.9 & 42,103 \\
Both models are length-stable dated API snapshots & ${+}0.97^{***}$ & (0.17) & 10.8 & 29,369 \\
\multicolumn{5}{l}{\textit{Interface}} \\
Fixed effects for each period between major page changes & ${+}2.61^{***}$ & (0.48) & 17.1 & 1,260,967 \\
Fixed effects for each period between any page changes & ${+}1.03^{***}$ & (0.31) & 17.1 & 1,260,967 \\
Seven days around each major page change dropped & ${+}1.82^{***}$ & (0.17) & 18.0 & 824,591 \\
Three days around each page change dropped & ${+}1.77^{***}$ & (0.15) & 17.6 & 810,702 \\
Traffic surge of 7--14 August 2024 dropped & ${+}1.56^{***}$ & (0.17) & 16.9 & 1,217,518 \\
\multicolumn{5}{l}{\textit{Time support and weighting}} \\
January to August 2024 only & ${+}1.91^{***}$ & (0.23) & 17.2 & 1,155,923 \\
Pairs judged over at least three months & ${+}1.54^{***}$ & (0.17) & 14.5 & 439,492 \\
Pairs judged over at least six months & ${+}1.10^{***}$ & (0.16) & 12.4 & 34,585 \\
Pair-by-month cells, equally weighted & ${+}1.45^{***}$ & (0.17) & 19.9 & 5,739 \\
Most active 1\% of judge identifiers dropped & ${+}1.65^{***}$ & (0.18) & 16.7 & 985,578 \\
\multicolumn{5}{l}{\textit{Pair strength}} \\
Weakest third of pairs & ${+}1.63^{***}$ & (0.17) & 20.4 & 420,399 \\
Middle third of pairs & ${+}1.76^{***}$ & (0.25) & 16.3 & 420,310 \\
Strongest third of pairs & ${+}1.75^{***}$ & (0.23) & 14.5 & 420,258 \\
\bottomrule
\end{tabularx}

\par\smallskip
\begin{minipage}{0.97\linewidth}\footnotesize
\textit{Notes.} Each row is a linear probability model of a ``both are bad'' vote on months since 1 May 2023 with model-pair fixed effects (in the interface rows, also fixed effects for each period between page changes). The pair-by-month row uses cell means. Standard errors clustered by pair and calendar week (pair and month for cells). Major page changes changed at least 50 lines of the battle page. Pair strength is the mean Bradley--Terry rating of the two models. $^{*}p<0.05$, $^{**}p<0.01$, $^{***}p<0.001$.
\end{minipage}
\end{table}

\subsection{People become pickier}

Table~\ref{tab:judges} turns to the judges. If the rise came only from a changing pool of voters, it would disappear when each voter is compared with themselves. It does not. The \WJJudges judge identifiers observed over at least a month make up \WJShareIds\% of identifiers but cast \WJShareVotes\% of the votes on frozen pairs. Among these identifiers, with identifier and pair fixed effects, the share of ``both are bad'' votes rises by \WJAbs points a month (SE \WJSE); among identifiers observed over at least three months it rises by \WJThreeAbs. Comparing the same identifier on the same pair, using the \JPN votes in the \JPGroups identifier-pair combinations judged more than a month apart (\JPJudges identifiers, \JPPairs pairs), gives \JPB (SE \JPSE), and \JPCtrlB with prompt-content controls. Controlling for the number of earlier votes leaves \WJLkB points a month, and dropping the most active 1\% of identifiers leaves \WJNoProlificB. The standard error is similar when clustering by identifier and week (\WJClJWSE). Hypothesis~\ref{h:judges}(a) holds: within the same identifier, and within the same identifier-pair combination, verdicts become harsher.

\begin{figure}[t]
\centering
\includegraphics[width=0.62\linewidth]{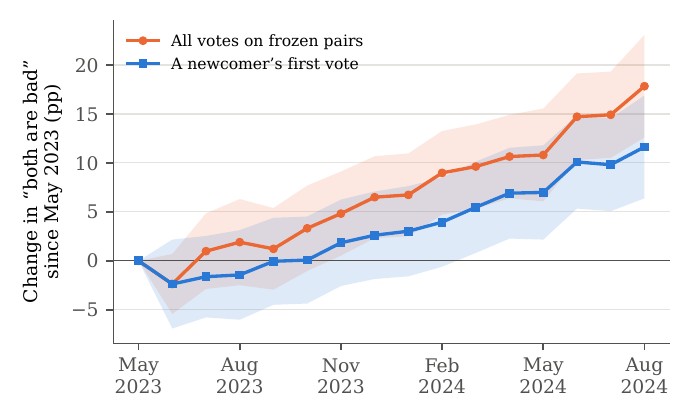}
\caption{Newcomers arrive pickier. Calendar-month effects (relative to May 2023) on ``both are bad'' within frozen model pairs, for all votes and for each judge identifier's first recorded vote by month of arrival. Shaded bands are 95\% confidence intervals, standard errors clustered by pair and week.}
\label{fig:newcomers}
\end{figure}

Arrival cohorts tell the same story. Using only each identifier's first recorded vote, voters who arrive later are harsher on the same frozen pairs: \NewB points per month of arrival date (SE \NewSE), \NewCtrlB with prompt-content and conversation controls, \NewEngB for English prompts and \NewNoSurgeB without the final week. Figure~\ref{fig:newcomers} shows the climb: by August 2024 a first recorded verdict on the same pair is \NewMonthLast points (SE \NewMonthLastSE) more likely to be ``both are bad'' than one in May 2023. A first recorded vote follows no earlier recorded vote on Arena, so the cohort pattern reflects what voters bring with them when they arrive, as Hypothesis~\ref{h:judges}(b) predicts.

Recorded experience goes with harsher verdicts too. In the same calendar month, votes cast after at least ten earlier votes are \VetAbs points more likely to be ``both are bad'' than earlier votes (SE \VetSE). Both patterns point to the judges: the same identifier becomes harsher over time, and later cohorts start harsher than earlier ones.

\begin{table}[t]
\centering
\caption{Who becomes pickier}
\label{tab:judges}
\small
\begin{tabularx}{\textwidth}{Xlrrr}
\toprule
Regressor (outcome: ``both are bad'', pp) & Per & Coef. & (SE) & Votes \\
\midrule
\multicolumn{5}{l}{\textit{Same identifier over time}} \\
Identifiers seen over $\geq$1 month (identifier and pair FE) & month & ${+}1.90^{***}$ & (0.13) & 365,074 \\
Identifiers seen over $\geq$3 months & month & ${+}1.74^{***}$ & (0.16) & 197,193 \\
Same identifier on the same pair, judged $\geq$1 month apart (identifier-by-pair FE) & month & ${+}2.30^{***}$ & (0.68) & 11,642 \\
\quad with prompt-content controls & month & ${+}2.24^{***}$ & (0.67) & 11,642 \\
Identifiers seen over $\geq$1 month, controlling for log(1 + earlier votes) & month & ${+}1.50^{***}$ & (0.13) & 365,074 \\
Identifiers seen over $\geq$1 month, most active 1\% dropped & month & ${+}1.83^{***}$ & (0.12) & 295,335 \\
\multicolumn{5}{l}{\textit{Arrival cohorts: first recorded vote}} \\
By month of arrival & month & ${+}1.25^{***}$ & (0.13) & 344,524 \\
\quad with prompt-content and conversation controls & month & ${+}1.15^{***}$ & (0.13) & 344,524 \\
\quad English prompts only & month & ${+}1.17^{***}$ & (0.13) & 202,692 \\
\quad traffic surge of 7--14 August 2024 dropped & month & ${+}1.17^{***}$ & (0.12) & 334,491 \\
\multicolumn{5}{l}{\textit{Experience, same calendar month}} \\
log(1 + earlier votes) & log unit & ${+}1.10^{***}$ & (0.14) & 1,260,967 \\
At least ten earlier votes & indicator & ${+}2.56^{***}$ & (0.31) & 1,260,967 \\
\multicolumn{5}{l}{\textit{Previous encounter}} \\
Better model in the previous battle (per 100 Elo) & 100 Elo & $-0.11$ & (0.09) & 916,214 \\
\quad with identifier fixed effects & 100 Elo & $-0.25^{**}$ & (0.09) & 916,214 \\
\bottomrule
\end{tabularx}

\par\smallskip
\begin{minipage}{0.97\linewidth}\footnotesize
\textit{Notes.} Outcome: a ``both are bad'' vote on a frozen pair, in percentage points. All rows include model-pair fixed effects or identifier-by-pair fixed effects. Same-identifier rows add identifier fixed effects; arrival-cohort rows use each identifier's first recorded vote; experience rows add calendar-month fixed effects; previous-encounter rows add week fixed effects and fixed effects for the number of earlier battles (1 to 29, and 30 or more), plus arrival-week fixed effects or identifier fixed effects. Standard errors clustered by pair and by week (arrival-cohort rows) or by identifier and by pair (other rows). $^{*}p<0.05$, $^{**}p<0.01$, $^{***}p<0.001$.
\end{minipage}
\end{table}

\subsection{No momentary contrast}

The last rows of Table~\ref{tab:judges} ask whether verdicts jump after a single encounter. A momentary contrast would make a voter harsher right after meeting a strong model. The data show no such effect. The rating of the better model in a voter's previous recorded battle is unrelated to their next verdict on a frozen pair (\PrevB points per 100 rating points, SE \PrevSE). With identifier fixed effects the coefficient is \PrevJB (SE \PrevJSE), so a strong previous model goes with a slightly milder next verdict, not a harsher one. Hypothesis~\ref{h:gradual} holds. The absence of a short-run contrast fits a reference that pools experience over time.

The dates on which new top-rated models first appeared on Arena show mixed co-movements (Appendix Table~\ref{tab:frontier}). The arrival of \texttt{gpt-4o-2024-05-13} coincides with a jump of \FrontGPTFourO points and that of \texttt{gpt-4-turbo-2024-04-09} with a fall of \FrontTurbo points; the largest change coincides with the traffic surge of 7 August 2024. The monthly path in Figure~\ref{fig:treadmill}(b) climbs in most months, including months without such arrivals. The average rating of the models in that month's battles is closely associated with the harsher verdicts (\PlatMeanAbs points per 100 rating points). It rises almost in a straight line with calendar time (correlation \CorrMeanEloMonth), so this association and the monthly trend describe the same steady climb.

\subsection{A shared drift that frozen references can remove}

\begin{figure}[t]
\centering
\includegraphics[width=0.8\linewidth]{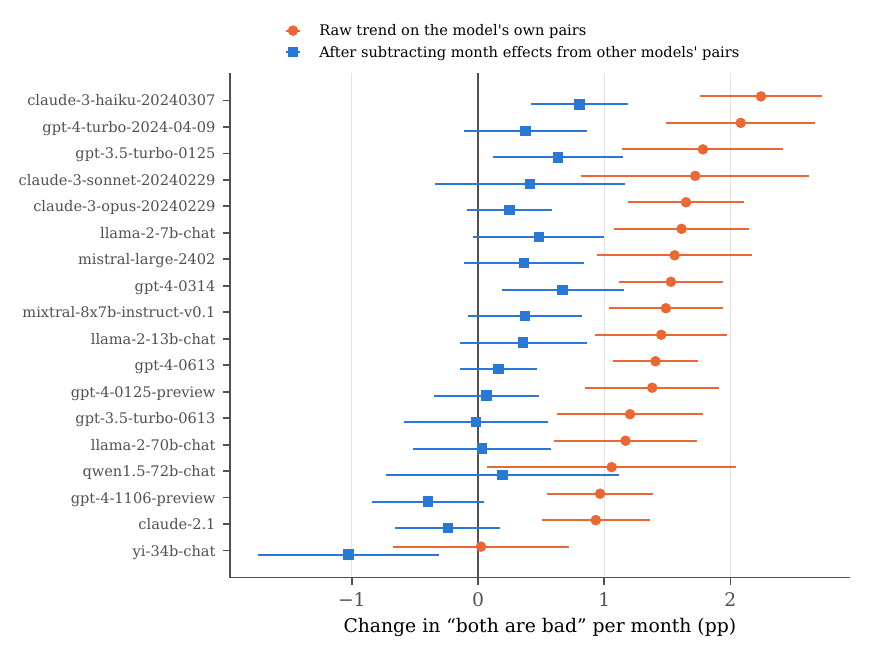}
\caption{A shared drift. For each frozen model with at least 10,000 votes over at least four months: the trend in ``both are bad'' on the model's own pairs (orange) and the trend after subtracting calendar-month effects estimated only from frozen pairs that do not involve the model (blue). Pair fixed effects; 95\% confidence intervals, standard errors clustered by pair and week.}
\label{fig:reanchor}
\end{figure}

The last test asks whether the drift is shared across models, and whether it can be removed out of sample (Figure~\ref{fig:reanchor}). Of the \RAModels frozen models with enough votes, \RARawPosLow show a significant rise in ``both are bad'' on their own pairs, with an average trend of \RAMeanRaw points a month. A platform that tracked the acceptance of each fixed model separately would conclude that \RARawPosLow of these models had become worse. After subtracting month effects estimated only from frozen pairs that do not involve the model, the average trend is \RAMeanAdj points a month; \RAAdjPos models keep a significantly positive trend and \RAAdjNeg a significantly negative one. The drift on any one model's pairs is predicted by the drift on the other models' pairs. The drift is therefore a shift common to the evaluations rather than a set of separate changes in individual models, and a reference set of frozen pairs measured in the same period is enough to remove it.

\section{Discussion}
\label{sec:discussion}

For more than a year, voters on Chatbot Arena judged unchanged AI models more harshly month after month. In 2024, when they chose between two answers, they chose the stronger model about as often at the end as at the start. What moved was how often neither answer was good enough, and it moved within the same voters, on the same pairs, and across cohorts of newcomers. Measured changes in prompts, in the serving of the models and in the interface do not account for it, and the drift on each model's pairs is predicted by the drift on the others'. On the platform as a whole, the models improved greatly while the share of answers rejected as ``both are bad'' ended near where it started.

\subsection{Theoretical implications}

\textbf{Evaluations that move while the system stands still.} Continuance research models expectations about a focal system and follows how they change with use of that system \citep{bhattacherjee2001understanding, bhattacherjee2004understanding, kim2005longitudinal}. Our evidence shows evaluations of fixed systems changing for reasons outside the systems. Verdicts became harsher within voters, and voters who arrived later were harsher from their first recorded vote, although nothing about the systems had changed. Adaptation-level theory accounts for this pattern with a reference formed by experience across a category \citep{helson1964adaptation, parducci1965category}: as the category improves, the standard against which a system is judged rises, so evaluations of a system that stands still decline. Expectation-confirmation models of fast-moving technologies can incorporate this account by letting the expectation term depend on exposure to other systems in the category, with boundary conditions such as how substitutable the systems are and how visible the category's progress is.

\textbf{A rater-side source of declining ratings.} Declining ratings for a fixed product are usually explained by changes in who rates \citep{li2008self, godes2012sequential, moe2012online}. Our design holds both the product and the voter identifier fixed and still finds decline, so the drift of each rater's standard is a separate mechanism. It implies that ratings of unchanged products should fall fastest in categories where quality improves fastest, a prediction that can be tested in any rating system.

\textbf{Absolute and relative judgments of AI.} The split between a stable ranking and a drifting bar is what work on comparative and absolute judgment predicts \citep{thurstone1927law, parducci1965category, stewart2005absolute}. It mirrors, in reverse, reputation inflation on online platforms, where the meaning of a fixed score drifts upward \citep{filippas2022reputation}. For AI, the meaning of ``acceptable'' drifted the other way.

\subsection{Practical implications}

\textbf{For AI providers.} Complaints that a model ``got worse'' can arise when nothing about the model changed. A provider that keeps a model fixed should expect evaluations of it to fall as the rest of the market improves: on Arena, the share of outright rejections of fixed pairs rose at an annualized historical rate of about \MainYear percentage points on a base of \MainMean\%. Separating a real change in a product from rising standards requires a fixed reference evaluated alongside the product over time.

\textbf{For evaluation platforms and leaderboards.} Absolute verdicts such as ``both are bad'' are not comparable across periods without adjustment. Platforms can keep frozen reference models in the evaluation mix, as Arena did by accident, and use them to re-anchor absolute measures, in the spirit of re-anchoring rating scales \citep{garg2021designing}. Our out-of-sample test shows that this removes the drift from models whose pairs were not used to estimate it. The result suggests a simple design principle for evaluation systems of fast-moving technologies: keep a reference set of fixed items in the evaluation stream, estimate period effects on that set, and report absolute measures net of those effects alongside the raw measures. Frozen models still meet new prompts; fixed prompt--answer pairs would hold the task fixed as well and so offer a stricter anchor. Asking for a forced choice and a separate acceptability judgment on every comparison would record both quantities directly. On Arena, the relative verdicts that drive the rankings showed no trend in 2024 while the absolute ones drifted. Not all drift should be removed: if users' needs have genuinely risen, a current measure of satisfaction should reflect that, so platforms gain most from reporting both a fixed-standard measure and a current one.

\textbf{For human feedback in training.} Preference data used to train models mix relative choices with absolute judgments of acceptability \citep{christiano2017deep, ouyang2022training}. Our results imply that absolute labels carry a time stamp: a response of a given quality that was accepted in 2023 would be more likely to be rejected by a comparable voter in 2024. Pipelines that pool absolute labels across periods, or use them to set a fixed quality threshold, mix different standards. Whether that helps, by tracking what users now want, or hurts, by adding noise to a fixed target, depends on the purpose of the label.

\subsection{Limitations and future work}

The first set of limits concerns who is observed. Arena's voters are a self-selected public interested in comparing AI, so the size of the drift in the general population is a question for future work, for example with panels that ask the same people to rate fixed answers at intervals. The judge identifier follows a device or network rather than a person, the log covers only voters who chose to vote, and the previous-battle test conditions on voters who returned. Later arrival cohorts may differ in who they are as well as in what they have experienced.

The second set concerns what the log records. It contains no prompt or answer text and no prompt key, so the high-frequency subset does not hold the mix of prompts fixed; text releases covering parts of the period would allow comparisons of verdicts on identical prompts and on the answers themselves. The interface tests rely on commit dates, which can differ from deployment dates, and the arrival dates of new models mark first appearances on Arena rather than measured exposure. The relative verdict is a pooled share, which can hide offsetting changes in individual pairs, and the intervals for the fourteen long-observed pairs treat pairs and votes as independent.

The third set concerns the mechanism. The data locate the change in the judges but do not record each voter's exposure to other AI systems. Because the average quality of the models on Arena rose almost in step with calendar time, the data also cannot separate it from other influences that grew over the same months, and the vote cannot separate a higher threshold from closer scrutiny. Linking verdicts to measured exposure, and using answers with known flaws to separate the two components, are the next steps. The most direct test is an experiment that exposes people to stronger or weaker AI on matched tasks and then has them rate the same saved answer pairs. Recording a forced choice and an acceptability judgment for each answer, with a delayed follow-up, would show both the size of the effect and how long it lasts. Finally, the period ends in August 2024. Whether standards keep rising at the same pace, or level off as improvement slows, can be tracked as new frozen models accumulate on evaluation platforms.

\section{Conclusion}
\label{sec:conclusion}

Organizations read human evaluations of AI as evidence about the systems being evaluated. This paper shows that the evaluations themselves move. Our three questions have clear answers. On Chatbot Arena, verdicts on unchanged models became steadily harsher from 2023 to 2024. The change sat in the absolute judgment rather than the relative one, and it appeared within raters and across arrival cohorts rather than only through a changing pool. And because the drift was shared across models, frozen references measured in the same period removed it. For a technology that improves as quickly as generative AI, a product that stands still is judged as declining, and evaluation systems that separate the product from the standard applied to it give a truer picture of both.

\bibliographystyle{plainnat}
\bibliography{references}

\appendix
\section{Model ratings}
\label{app:ratings}

Model strength is a Bradley--Terry rating \citep{bradley1952rank} estimated once on all \NWindow battles in the window, using the minorization--maximization algorithm of \citet{hunter2004mm}. Ordinary ties and ``both are bad'' votes each count as half a win for both sides. Ratings are reported on the Elo scale (400 times the base-10 logarithm of the Bradley--Terry strength) and normalized so that \texttt{gpt-4-0613} has a rating of 1000. A difference of $\Delta$ rating points implies an expected score of $1/(1+10^{-\Delta/400})$ for the stronger model, counting a tie as half a win. Ratings enter the analysis only to define the stronger model of a pair, the strength of a pair, and the monthly platform series. Two alternatives give the same results for the relative verdict (Section~\ref{sec:results}): ratings estimated on decisive votes only, which correlate at \CorrEloNoTies with the main ratings, and cross-fitted ratings estimated on alternate calendar weeks. The expected score in Section~\ref{sec:results} averages the expected score over all pairs of a model appearance in August 2024 and a model appearance in May 2023.

\section{Model classification}
\label{app:models}

Table~\ref{tab:models} lists every model in the window with its class, the months of its first and last battle, and its number of appearances. Each model is assigned to its class by an explicit list; $^{\dagger}$ marks the dated OpenAI and Anthropic snapshots, the strictest tier.

{\small
\setlength{\LTcapwidth}{\linewidth}
\begin{center}
\captionof{table}{Models by class}
\label{tab:models}
\end{center}
\begin{longtable}{lccr}
\toprule
Model & First & Last & Appearances \\
\midrule
\endhead
\multicolumn{4}{l}{\textit{Dated API snapshots} (33)} \\
\texttt{gpt-4-0314}$^{\dagger}$ & 2023-05 & 2024-06 & 59,852 \\
\texttt{gpt-3.5-turbo-0314}$^{\dagger}$ & 2023-05 & 2023-06 & 5,841 \\
\texttt{gpt-4-0613}$^{\dagger}$ & 2023-06 & 2024-08 & 96,284 \\
\texttt{gpt-3.5-turbo-0613}$^{\dagger}$ & 2023-06 & 2024-03 & 40,549 \\
\texttt{claude-2.0}$^{\dagger}$ & 2023-07 & 2024-03 & 13,323 \\
\texttt{gpt-4-1106-preview}$^{\dagger}$ & 2023-11 & 2024-08 & 98,857 \\
\texttt{gpt-3.5-turbo-1106}$^{\dagger}$ & 2023-11 & 2024-02 & 17,693 \\
\texttt{claude-2.1}$^{\dagger}$ & 2023-11 & 2024-04 & 39,426 \\
\texttt{gpt-4-0125-preview}$^{\dagger}$ & 2024-01 & 2024-08 & 93,166 \\
\texttt{gpt-3.5-turbo-0125}$^{\dagger}$ & 2024-02 & 2024-07 & 74,688 \\
\texttt{mistral-large-2402} & 2024-02 & 2024-07 & 70,488 \\
\texttt{claude-3-sonnet-20240229}$^{\dagger}$ & 2024-03 & 2024-07 & 123,261 \\
\texttt{claude-3-opus-20240229}$^{\dagger}$ & 2024-03 & 2024-08 & 168,103 \\
\texttt{claude-3-haiku-20240307}$^{\dagger}$ & 2024-03 & 2024-08 & 119,736 \\
\texttt{gemini-1.5-pro-api-0409-preview} & 2024-04 & 2024-05 & 62,366 \\
\texttt{reka-flash-21b-20240226} & 2024-04 & 2024-06 & 29,139 \\
\texttt{gpt-4-turbo-2024-04-09}$^{\dagger}$ & 2024-04 & 2024-08 & 93,756 \\
\texttt{gemini-1.5-flash-api-0514} & 2024-05 & 2024-08 & 59,087 \\
\texttt{gemini-1.5-pro-api-0514} & 2024-05 & 2024-08 & 73,065 \\
\texttt{reka-core-20240501} & 2024-05 & 2024-07 & 70,632 \\
\texttt{glm-4-0116} & 2024-05 & 2024-05 & 8,207 \\
\texttt{qwen-max-0428} & 2024-05 & 2024-06 & 30,255 \\
\texttt{gpt-4o-2024-05-13}$^{\dagger}$ & 2024-05 & 2024-08 & 81,771 \\
\texttt{claude-3-5-sonnet-20240620}$^{\dagger}$ & 2024-06 & 2024-08 & 49,530 \\
\texttt{glm-4-0520} & 2024-06 & 2024-07 & 10,451 \\
\texttt{reka-flash-preview-20240611} & 2024-06 & 2024-07 & 21,252 \\
\texttt{gpt-4o-mini-2024-07-18}$^{\dagger}$ & 2024-07 & 2024-08 & 20,302 \\
\texttt{deepseek-v2-api-0628} & 2024-07 & 2024-08 & 16,912 \\
\texttt{mistral-large-2407} & 2024-07 & 2024-08 & 12,026 \\
\texttt{gemini-1.5-pro-exp-0801} & 2024-07 & 2024-08 & 21,099 \\
\texttt{reka-core-20240722} & 2024-07 & 2024-08 & 7,346 \\
\texttt{reka-flash-20240722} & 2024-07 & 2024-08 & 7,480 \\
\texttt{gpt-4o-2024-08-06}$^{\dagger}$ & 2024-08 & 2024-08 & 10,235 \\
\multicolumn{4}{l}{\textit{Fixed open weights} (73)} \\
\texttt{RWKV-4-Raven-14B} & 2023-05 & 2023-08 & 5,084 \\
\texttt{mpt-7b-chat} & 2023-05 & 2023-08 & 4,161 \\
\texttt{oasst-pythia-12b} & 2023-05 & 2023-08 & 5,566 \\
\texttt{llama-13b} & 2023-05 & 2023-07 & 1,708 \\
\texttt{koala-13b} & 2023-05 & 2023-07 & 5,908 \\
\texttt{fastchat-t5-3b} & 2023-05 & 2023-08 & 4,388 \\
\texttt{dolly-v2-12b} & 2023-05 & 2023-07 & 2,784 \\
\texttt{stablelm-tuned-alpha-7b} & 2023-05 & 2023-06 & 2,653 \\
\texttt{alpaca-13b} & 2023-05 & 2023-08 & 5,256 \\
\texttt{chatglm-6b} & 2023-05 & 2023-08 & 4,418 \\
\texttt{wizardlm-13b} & 2023-06 & 2023-11 & 7,545 \\
\texttt{mpt-30b-chat} & 2023-06 & 2023-09 & 2,774 \\
\texttt{guanaco-33b} & 2023-06 & 2023-08 & 3,152 \\
\texttt{gpt4all-13b-snoozy} & 2023-06 & 2023-07 & 1,888 \\
\texttt{llama-2-7b-chat} & 2023-07 & 2024-05 & 15,565 \\
\texttt{llama-2-13b-chat} & 2023-07 & 2024-05 & 21,207 \\
\texttt{codellama-34b-instruct} & 2023-08 & 2024-02 & 7,972 \\
\texttt{llama-2-70b-chat} & 2023-08 & 2024-05 & 42,717 \\
\texttt{chatglm2-6b} & 2023-08 & 2023-11 & 2,855 \\
\texttt{wizardlm-70b} & 2023-09 & 2024-02 & 8,784 \\
\texttt{mistral-7b-instruct} & 2023-09 & 2024-02 & 9,549 \\
\texttt{falcon-180b-chat} & 2023-09 & 2024-01 & 1,381 \\
\texttt{qwen-14b-chat} & 2023-10 & 2024-02 & 5,252 \\
\texttt{zephyr-7b-beta} & 2023-10 & 2024-02 & 11,826 \\
\texttt{zephyr-7b-alpha} & 2023-10 & 2023-11 & 1,888 \\
\texttt{openhermes-2.5-mistral-7b} & 2023-11 & 2024-03 & 5,242 \\
\texttt{starling-lm-7b-alpha} & 2023-11 & 2024-04 & 10,865 \\
\texttt{tulu-2-dpo-70b} & 2023-11 & 2024-03 & 6,890 \\
\texttt{chatglm3-6b} & 2023-11 & 2024-02 & 4,903 \\
\texttt{yi-34b-chat} & 2023-11 & 2024-05 & 17,438 \\
\texttt{openchat-3.5} & 2023-11 & 2024-02 & 8,381 \\
\texttt{solar-10.7b-instruct-v1.0} & 2023-12 & 2024-01 & 4,425 \\
\texttt{mixtral-8x7b-instruct-v0.1} & 2023-12 & 2024-08 & 81,858 \\
\texttt{dolphin-2.2.1-mistral-7b} & 2023-12 & 2024-01 & 1,764 \\
\texttt{llama2-70b-steerlm-chat} & 2024-01 & 2024-02 & 3,717 \\
\texttt{deepseek-llm-67b-chat} & 2024-01 & 2024-02 & 5,139 \\
\texttt{stripedhyena-nous-7b} & 2024-01 & 2024-02 & 5,449 \\
\texttt{gemma-2b-it} & 2024-02 & 2024-04 & 5,263 \\
\texttt{mistral-7b-instruct-v0.2} & 2024-02 & 2024-05 & 21,598 \\
\texttt{qwen1.5-72b-chat} & 2024-02 & 2024-06 & 44,260 \\
\texttt{codellama-70b-instruct} & 2024-02 & 2024-04 & 1,312 \\
\texttt{qwen1.5-4b-chat} & 2024-02 & 2024-04 & 8,404 \\
\texttt{nous-hermes-2-mixtral-8x7b-dpo} & 2024-02 & 2024-03 & 3,936 \\
\texttt{gemma-7b-it} & 2024-02 & 2024-04 & 9,777 \\
\texttt{openchat-3.5-0106} & 2024-02 & 2024-04 & 14,023 \\
\texttt{qwen1.5-7b-chat} & 2024-02 & 2024-05 & 5,288 \\
\texttt{starling-lm-7b-beta} & 2024-03 & 2024-05 & 18,047 \\
\texttt{olmo-7b-instruct} & 2024-03 & 2024-04 & 6,942 \\
\texttt{qwen1.5-32b-chat} & 2024-04 & 2024-06 & 25,887 \\
\texttt{gemma-1.1-7b-it} & 2024-04 & 2024-06 & 27,446 \\
\texttt{qwen1.5-14b-chat} & 2024-04 & 2024-05 & 21,328 \\
\texttt{zephyr-orpo-141b-A35b-v0.1} & 2024-04 & 2024-05 & 5,408 \\
\texttt{phi-3-mini-128k-instruct} & 2024-04 & 2024-05 & 24,320 \\
\texttt{llama-3-70b-instruct} & 2024-04 & 2024-08 & 182,232 \\
\texttt{llama-3-8b-instruct} & 2024-04 & 2024-08 & 117,970 \\
\texttt{mixtral-8x22b-instruct-v0.1} & 2024-04 & 2024-08 & 55,401 \\
\texttt{gemma-1.1-2b-it} & 2024-04 & 2024-06 & 12,678 \\
\texttt{snowflake-arctic-instruct} & 2024-04 & 2024-06 & 37,878 \\
\texttt{yi-1.5-34b-chat} & 2024-05 & 2024-07 & 26,292 \\
\texttt{qwen1.5-110b-chat} & 2024-05 & 2024-06 & 31,377 \\
\texttt{phi-3-small-8k-instruct} & 2024-05 & 2024-07 & 19,311 \\
\texttt{phi-3-mini-4k-instruct} & 2024-05 & 2024-07 & 22,351 \\
\texttt{phi-3-medium-4k-instruct} & 2024-05 & 2024-08 & 23,444 \\
\texttt{qwen2-72b-instruct} & 2024-05 & 2024-08 & 35,809 \\
\texttt{gemma-2-27b-it} & 2024-06 & 2024-08 & 28,957 \\
\texttt{gemma-2-9b-it} & 2024-06 & 2024-08 & 26,281 \\
\texttt{nemotron-4-340b-instruct} & 2024-06 & 2024-07 & 21,417 \\
\texttt{gemma-2-2b-it} & 2024-07 & 2024-08 & 11,004 \\
\texttt{phi-3-mini-4k-instruct-june-2024} & 2024-07 & 2024-08 & 10,984 \\
\texttt{llama-3.1-405b-instruct} & 2024-07 & 2024-08 & 19,768 \\
\texttt{llama-3.1-70b-instruct} & 2024-07 & 2024-08 & 14,277 \\
\texttt{athene-70b-0725} & 2024-07 & 2024-08 & 12,046 \\
\texttt{llama-3.1-8b-instruct} & 2024-07 & 2024-08 & 13,254 \\
\multicolumn{4}{l}{\textit{Aliases that may change (excluded from frozen pairs)} (23)} \\
\texttt{vicuna-7b} & 2023-05 & 2023-11 & 7,319 \\
\texttt{claude-1} & 2023-05 & 2024-03 & 21,608 \\
\texttt{claude-instant-1} & 2023-05 & 2024-03 & 21,401 \\
\texttt{vicuna-13b} & 2023-05 & 2024-05 & 18,955 \\
\texttt{palm-2} & 2023-05 & 2023-12 & 9,145 \\
\texttt{vicuna-33b} & 2023-06 & 2024-04 & 24,172 \\
\texttt{pplx-7b-online} & 2023-11 & 2024-02 & 6,573 \\
\texttt{pplx-70b-online} & 2023-11 & 2024-02 & 7,183 \\
\texttt{mistral-medium} & 2023-12 & 2024-05 & 37,887 \\
\texttt{gemini-pro} & 2023-12 & 2024-01 & 6,784 \\
\texttt{bard-jan-24-gemini-pro} & 2024-01 & 2024-03 & 12,258 \\
\texttt{gemini-pro-dev-api} & 2024-01 & 2024-04 & 19,903 \\
\texttt{mistral-next} & 2024-02 & 2024-03 & 12,948 \\
\texttt{dbrx-instruct-preview} & 2024-03 & 2024-06 & 37,328 \\
\texttt{command-r} & 2024-03 & 2024-07 & 62,448 \\
\texttt{reka-flash-21b-20240226-online} & 2024-04 & 2024-05 & 18,320 \\
\texttt{command-r-plus} & 2024-04 & 2024-07 & 89,811 \\
\texttt{gemini-advanced-0514} & 2024-05 & 2024-08 & 55,117 \\
\texttt{yi-large} & 2024-05 & 2024-07 & 17,305 \\
\texttt{yi-large-preview} & 2024-05 & 2024-07 & 54,992 \\
\texttt{deepseek-coder-v2} & 2024-06 & 2024-07 & 16,293 \\
\texttt{deepseek-coder-v2-0724} & 2024-07 & 2024-08 & 4,186 \\
\texttt{chatgpt-4o-latest} & 2024-08 & 2024-08 & 15,268 \\
\bottomrule
\end{longtable}

\addtocounter{table}{-1}
}

\section{Serving stability}
\label{app:serving}

For each frozen model active for at least three months with at least 2,000 single-turn answers, Table~\ref{tab:serving} reports the trend in the logarithm of its own answer length, in percent per month, from a regression with model-by-opponent fixed effects and controls for prompt length and English. A model is length-stable if the trend is below 0.5\% a month in absolute value, or below 1\% a month and not significant at the 5\% level. Of the \NServingChecked models checked, \NStableModels are length-stable. Answer length is a diagnostic of serving changes that alter what a model writes; it complements the classification by version and the separate estimates for dated OpenAI and Anthropic snapshots and open-weight models.

{\small
\begin{center}
\captionof{table}{Trend in each frozen model's answer length}
\label{tab:serving}
\end{center}
\begin{longtable}{llrrrrc}
\toprule
Model & Class & Months & Answers & \%/month & (SE) & Stable \\
\midrule
\endhead
\texttt{chatglm-6b} & open weights & 3.4 & 2,885 & $-22.7$ & (5.2) &  \\
\texttt{alpaca-13b} & open weights & 3.7 & 3,360 & $-15.4$ & (4.0) &  \\
\texttt{oasst-pythia-12b} & open weights & 3.4 & 3,609 & $-14.2$ & (5.9) &  \\
\texttt{fastchat-t5-3b} & open weights & 3.6 & 2,894 & $-7.8$ & (4.6) &  \\
\texttt{gpt-4-0125-preview} & snapshot & 6.6 & 66,961 & $-4.0$ & (1.0) &  \\
\texttt{qwen1.5-72b-chat} & open weights & 4.2 & 31,750 & $-3.9$ & (1.6) &  \\
\texttt{claude-3-sonnet-20240229} & snapshot & 4.5 & 88,086 & $-3.8$ & (1.4) &  \\
\texttt{claude-3-opus-20240229} & snapshot & 5.4 & 121,894 & $-3.2$ & (1.1) &  \\
\texttt{mistral-7b-instruct-v0.2} & open weights & 3.1 & 15,654 & $-1.7$ & (1.4) &  \\
\texttt{gpt-3.5-turbo-0125} & snapshot & 5.9 & 54,813 & $-1.6$ & (0.9) &  \\
\texttt{mistral-large-2402} & snapshot & 4.9 & 51,269 & $-1.2$ & (0.7) &  \\
\texttt{claude-3-haiku-20240307} & snapshot & 5.0 & 82,487 & $-1.1$ & (1.1) &  \\
\texttt{claude-2.1} & snapshot & 4.9 & 25,030 & $-0.4$ & (0.8) & yes \\
\texttt{mixtral-8x7b-instruct-v0.1} & open weights & 7.9 & 56,986 & $-0.1$ & (0.7) & yes \\
\texttt{gpt-4-0314} & snapshot & 13.1 & 42,817 & $-0.1$ & (0.9) & yes \\
\texttt{gpt-4-0613} & snapshot & 13.6 & 69,204 & ${+}0.1$ & (0.7) & yes \\
\texttt{gpt-4o-2024-05-13} & snapshot & 3.1 & 57,877 & ${+}0.2$ & (1.3) & yes \\
\texttt{qwen-14b-chat} & open weights & 3.8 & 3,413 & ${+}0.6$ & (2.2) & yes \\
\texttt{llama-2-70b-chat} & open weights & 8.8 & 26,811 & ${+}0.8$ & (1.4) & yes \\
\texttt{gpt-4-1106-preview} & snapshot & 9.3 & 68,217 & ${+}1.0$ & (1.0) & yes \\
\texttt{mistral-7b-instruct} & open weights & 5.0 & 7,007 & ${+}1.3$ & (0.8) &  \\
\texttt{gpt-4-turbo-2024-04-09} & snapshot & 4.2 & 70,443 & ${+}2.1$ & (0.8) &  \\
\texttt{yi-34b-chat} & open weights & 5.9 & 12,327 & ${+}2.2$ & (1.8) &  \\
\texttt{llama-2-13b-chat} & open weights & 10.1 & 13,848 & ${+}2.2$ & (1.0) &  \\
\texttt{claude-2.0} & snapshot & 7.5 & 8,134 & ${+}2.3$ & (0.9) &  \\
\texttt{tulu-2-dpo-70b} & open weights & 3.6 & 3,792 & ${+}2.9$ & (1.8) &  \\
\texttt{mixtral-8x22b-instruct-v0.1} & open weights & 3.9 & 42,583 & ${+}3.2$ & (0.8) &  \\
\texttt{zephyr-7b-beta} & open weights & 3.7 & 8,671 & ${+}3.5$ & (0.7) &  \\
\texttt{llama-2-7b-chat} & open weights & 9.5 & 11,325 & ${+}3.8$ & (1.1) &  \\
\texttt{gpt-3.5-turbo-0613} & snapshot & 8.6 & 23,054 & ${+}3.8$ & (0.9) &  \\
\texttt{gpt-3.5-turbo-1106} & snapshot & 3.6 & 9,366 & ${+}4.1$ & (1.6) &  \\
\texttt{wizardlm-70b} & open weights & 5.7 & 4,829 & ${+}4.8$ & (1.2) &  \\
\texttt{llama-3-8b-instruct} & open weights & 3.9 & 92,067 & ${+}5.2$ & (1.2) &  \\
\texttt{openchat-3.5} & open weights & 3.2 & 6,197 & ${+}5.6$ & (2.2) &  \\
\texttt{starling-lm-7b-alpha} & open weights & 4.1 & 8,340 & ${+}5.9$ & (1.2) &  \\
\texttt{llama-3-70b-instruct} & open weights & 3.9 & 140,523 & ${+}5.9$ & (1.0) &  \\
\texttt{wizardlm-13b} & open weights & 5.6 & 4,872 & ${+}8.3$ & (2.2) &  \\
\texttt{openhermes-2.5-mistral-7b} & open weights & 3.1 & 3,683 & ${+}8.4$ & (2.3) &  \\
\texttt{codellama-34b-instruct} & open weights & 5.9 & 4,939 & ${+}9.0$ & (1.5) &  \\
\texttt{RWKV-4-Raven-14B} & open weights & 3.2 & 3,150 & ${+}17.8$ & (5.4) &  \\
\bottomrule
\end{longtable}

\addtocounter{table}{-1}
}

\section{Prompt languages and interface changes}
\label{app:language}

Table~\ref{tab:language} repeats the main specification separately for each prompt language with at least 20,000 votes on frozen pairs. The trend is positive and significant in every language, ranging from \LangMin to \LangMax points a month among the named languages. Table~\ref{tab:ui} lists the major changes to the anonymous battle page, those that changed at least 50 lines of its source file, with the commit, its description and the estimated jump in ``both are bad'' votes on frozen pairs at the commit date. Most jumps are small and insignificant, the significant ones are negative, and jumps of similar size appear at placebo dates.

\begin{table}[H]
\centering
\caption{Change in ``both are bad'' per month within frozen pairs, by prompt language}
\label{tab:language}
\begin{tabular}{lrrrr}
\toprule
Prompt language & pp/month & (SE) & Mean (\%) & Votes \\
\midrule
English & ${+}1.53$ & (0.15) & 17.2 & 716,620 \\
Chinese & ${+}2.38$ & (0.33) & 18.6 & 132,682 \\
Russian & ${+}1.78$ & (0.31) & 19.0 & 113,583 \\
unknown & ${+}1.72$ & (0.70) & 10.5 & 69,549 \\
German & ${+}1.59$ & (0.33) & 16.4 & 35,544 \\
Korean & ${+}3.75$ & (0.67) & 15.8 & 31,229 \\
Japanese & ${+}1.52$ & (0.43) & 16.0 & 25,218 \\
Spanish & ${+}1.74$ & (0.47) & 15.4 & 21,987 \\
\bottomrule
\end{tabular}

\par\smallskip
\begin{minipage}{0.8\linewidth}\footnotesize
\textit{Notes.} Languages with at least 20,000 votes on frozen pairs; ``unknown'' is the language label in the log. Model-pair fixed effects; standard errors clustered by pair and week.
\end{minipage}
\end{table}

\begin{table}[H]
\centering
\caption{Jumps in ``both are bad'' at major changes to the anonymous battle page}
\label{tab:ui}
\small
\begin{tabularx}{\textwidth}{llrXrr}
\toprule
Date & Commit & Lines & Description & Jump (pp) & (SE) \\
\midrule
2023-05-08 & \texttt{c1985314} & 74 & Improve docs and format code & $-3.09$ & (1.77) \\
2023-06-14 & \texttt{07abdb08} & 58 & Misc improvements to the gradio web server & $-2.06$ & (2.60) \\
2023-06-24 & \texttt{78fa7b4e} & 50 & Update gradio version to 3.35 & ${+}2.88$ & (2.33) \\
2023-11-12 & \texttt{e0b351a4} & 246 & UI and model change & $-1.91$ & (0.99) \\
2023-12-09 & \texttt{d82f7ec3} & 77 & Update UI and new models & ${+}0.76$ & (1.02) \\
2023-12-24 & \texttt{a28563b9} & 110 & Add new models (Perplexity, gemini) and Separate GPT versions & ${+}0.25$ & (0.92) \\
2024-02-05 & \texttt{22642048} & 115 & code update & $-0.30$ & (0.72) \\
2024-02-09 & \texttt{3f61c6e6} & 198 & Upgrade gradio to 4.17 & ${+}1.54$ & (0.80) \\
2024-04-15 & \texttt{50956158} & 346 & Code update & $-1.74$ & (0.35) \\
2024-05-10 & \texttt{e395370d} & 58 & Add Multimodal Textbox & $-2.26$ & (0.98) \\
2024-06-30 & \texttt{be5db521} & 240 & Merge operation and multimodal arena & $-0.43$ & (0.56) \\
\bottomrule
\end{tabularx}

\par\smallskip
\begin{minipage}{0.97\linewidth}\footnotesize
\textit{Notes.} Commits that changed at least 50 lines of the source file of the anonymous battle page in the public repository of the software that runs Chatbot Arena \citep{fastchat}, with the commit message as description. Each row is a local linear regression of a ``both are bad'' vote on an indicator for the time after the commit, with separate linear trends on each side and pair fixed effects estimated within the window, using votes within 21 days on each side (7 days before the first commit, which falls one week into the window). Standard errors clustered by day. Jumps estimated in the same way at \PlaceboN placebo dates, every seventh day at least ten days from a major change, average \PlaceboMeanAbs points in absolute value, and \PlaceboShareSig\% of them are significant at the 5\% level, against \UIShareSig\% at the major changes.
\end{minipage}
\end{table}

\section{Arrivals of new top-rated models}
\label{app:frontier}

Table~\ref{tab:frontier} estimates the change in ``both are bad'' votes on frozen pairs at each arrival of a model rated above every model that arrived before it (among models with at least 2,000 appearances), using the same local regression as Table~\ref{tab:ui}. The jumps have both signs: one large positive jump at the arrival of \texttt{gpt-4o-2024-05-13}, one negative jump at the arrival of \texttt{gpt-4-turbo-2024-04-09}, and two small changes. The largest, on 7 August 2024, coincides with a surge in traffic in the last week of the log, when daily battles rose from about \SurgeBefore to \SurgeAfter; the main estimate without that week is \NoSurgeB points a month.

\begin{table}[H]
\centering
\caption{Jumps in ``both are bad'' at arrivals of a new top-rated model}
\label{tab:frontier}
\begin{tabular}{llrrr}
\toprule
Date & New top-rated model & Jump (pp) & (SE) & Votes ($\pm$21 days) \\
\midrule
2023-11-07 & \texttt{gpt-4-1106-preview} & ${+}1.26$ & (1.56) & 26,931 \\
2024-04-10 & \texttt{gpt-4-turbo-2024-04-09} & $-1.29$ & (0.63) & 311,723 \\
2024-05-14 & \texttt{gpt-4o-2024-05-13} & ${+}4.95$ & (0.91) & 427,307 \\
2024-07-18 & \texttt{gemini-1.5-pro-exp-0801} & ${+}1.09$ & (1.21) & 199,469 \\
2024-08-07 & \texttt{chatgpt-4o-latest} & ${+}8.70$ & (0.46) & 147,034 \\
\bottomrule
\end{tabular}

\par\smallskip
\begin{minipage}{0.8\linewidth}\footnotesize
\textit{Notes.} Day after the model first appeared in blind battles. Local linear regressions of a ``both are bad'' vote within 21 days of the arrival (7 days after the last arrival, where the log ends), with separate linear trends on each side and pair fixed effects; standard errors clustered by day.
\end{minipage}
\end{table}

\section{Coefficient-stability sensitivity}
\label{app:oster}

Table~\ref{tab:oster} reports the coefficient-stability calculations of \citet{oster2019unobservable} for the main trend. The uncontrolled estimate is \MainB points a month with within-pair $R^2$ \OsterRZero; with prompt-content and conversation controls it is \CtrlConvB with within-pair $R^2$ \OsterROne. Under proportional selection, $\delta$ is the ratio of selection on unobserved to selection on observed controls that would bring the trend to zero, given an assumed maximum $R^2$ that a full set of controls could reach; the last column is the bias-adjusted trend when $\delta = 1$. The calculation uses the linear approximation in Oster's paper on the within-pair (pair-demeaned) data. The rows for prompt content only exclude the conversation controls, which can depend on earlier answers. The outcome is an individual binary vote, so even a complete set of prompt controls would leave most of its variation unexplained; the table shows how the conclusion changes as the assumed maximum rises.

\begin{table}[H]
\centering
\caption{Coefficient-stability sensitivity for the main trend}
\label{tab:oster}
\begin{tabular}{llrr}
\toprule
Observed controls & Maximum $R^2$ & $\delta$ for zero trend & Trend at $\delta = 1$ (pp/month) \\
\midrule
Prompt content and conversation & 0.026 (1.3 $\times$ controlled $R^2$) & 35.9 & ${+}1.53$ \\
 & 0.040 (2 $\times$ controlled $R^2$) & 10.8 & ${+}1.42$ \\
 & 0.050 & 7.1 & ${+}1.35$ \\
 & 0.100 & 2.7 & ${+}0.98$ \\
Prompt content only & 0.022 (1.3 $\times$ controlled $R^2$) & 67.5 & ${+}1.61$ \\
 & 0.035 (2 $\times$ controlled $R^2$) & 20.2 & ${+}1.55$ \\
 & 0.050 & 10.7 & ${+}1.48$ \\
 & 0.100 & 4.2 & ${+}1.25$ \\
\bottomrule
\end{tabular}

\par\smallskip
\begin{minipage}{0.85\linewidth}\footnotesize
\textit{Notes.} Within-pair linear probability model of a ``both are bad'' vote on months since 1 May 2023. Controls: 20 prompt-language indicators, code, math and seven difficulty tags (prompt content), plus multi-turn and prompt length (conversation).
\end{minipage}
\end{table}

\end{document}